\documentclass[aps,prd,twocolumn,showpacs,byrevtex]{revtex4-1}
\usepackage{times}
\usepackage{amsmath}
\usepackage{graphicx}
\usepackage{epstopdf}
\usepackage{color}
\usepackage{multirow}
\usepackage{setspace}
\usepackage{overpic}
\usepackage{amssymb}
\newcommand{\pp}{\pi^+\pi^-}
\newcommand{\kk}{K^+ K^-}
\newcommand{\LL}{\ell^+ \ell^-}
\newcommand{\EE}{e^+e^-}
\newcommand{\KK}{K^{0}_S K^{0}_L}
\newcommand{\ks}{K_{S}^{0}}
\newcommand{\pipipi}{\pi^+ \pi^- \pi^0}
\newcommand{\pipi}{\pi^+ \pi^-}
\newcommand{\MM}{\mu^+\mu^-}

\newcommand{\jpsi}{J/\psi}
\newcommand{\phijpsi}{\phi \jpsi}
\def\Journal#1#2#3#4{{#1} {\bf #2}, #3 (#4)}
\def\IJMP{Int. J. Mod. Phys. A}

\def\NIMA{Nucl. Instrum. Methods A}
\def\NPB{Nucl. Phys. B}
\def\PLB{Phys. Lett. B}
\def\PRL{Phys. Rev. Lett.}
\def\PRD{Phys. Rev. D}

\def\EPJC{Eur. Phys. J. C}
\def\HEPNP{HEP \& NP}
\def\CTP{Commun. Theor. Phys.}
\def\JPG{J. Phys. G}
\def\CPC{Chin. Phys. C}

\begin{document}

\graphicspath{{figure/}}
\DeclareGraphicsExtensions{.eps,.png,.ps}

\title{\quad\\[0.0cm] \boldmath
Search for the $Y(4140)$ via $\EE \to \gamma \phijpsi$ at
$\sqrt{s}=$ 4.23, 4.26 and 4.36~GeV}

\author{
  \begin{small}
    \begin{center}
      M.~Ablikim$^{1}$, M.~N.~Achasov$^{8,a}$, X.~C.~Ai$^{1}$,
      O.~Albayrak$^{4}$, M.~Albrecht$^{3}$, D.~J.~Ambrose$^{43}$,
      A.~Amoroso$^{47A,47C}$, F.~F.~An$^{1}$, Q.~An$^{44}$,
      J.~Z.~Bai$^{1}$
% ...
      \\
      \vspace{0.2cm}
      (BESIII Collaboration)\\
      \vspace{0.2cm} {\it
        $^{1}$ Institute of High Energy Physics, Beijing 100049, People's Republic of China\\
% ...
$^{3}$ Bochum Ruhr-University, D-44780 Bochum, Germany\\
        $^{4}$ Carnegie Mellon University, Pittsburgh, Pennsylvania 15213, USA\\
% ...
$^{8}$ G.I. Budker Institute of Nuclear Physics SB RAS (BINP), Novosibirsk 630090, Russia\\
% ...
$^{43}$ University of Rochester, Rochester, New York 14627, USA\\
        $^{44}$ University of Science and Technology of China, Hefei 230026, People's Republic of China\\
% ...
$^{47}$ (A)University of Turin, I-10125, Turin, Italy; (B)University of Eastern Piedmont, I-15121, Alessandria, Italy; (C)INFN, I-10125, Turin, Italy\\
% ...
$^{a}$ Also at the Novosibirsk State University, Novosibirsk, 630090, Russia\\
% ...
      }\end{center}
    \vspace{0.4cm}
  \end{small}
}

\affiliation{}

\vspace{0.4cm}
\date{\today}

\begin{spacing}{1.0}

\begin{abstract}

Using data samples collected at center-of-mass energies $\sqrt{s} =
4.23$, 4.26, and 4.36~GeV with the BESIII detector operating at
the BEPCII storage ring, we search for the production of the charmoniumlike
state $Y(4140)$ through a radiative transition
followed by its decay to $\phi\jpsi$. No significant signal is
observed and upper limits on $\sigma[\EE \rightarrow \gamma
Y(4140)] \cdot \mathcal{B}(Y(4140)\rightarrow \phijpsi)$ at the $90\%$
confidence level are estimated as 0.35, 0.28, and
0.33~pb at $\sqrt{s} = 4.23$, 4.26, and 4.36~GeV, respectively.

\end{abstract}

\pacs{14.40.Rt, 13.66.Bc, 14.40.Pq, 13.20.Gd}

\maketitle


\section{Introduction}

The CDF experiment first reported
evidence for a new state called
 $Y(4140)$ in the
decay $B^{+} \rightarrow \phijpsi K^{+}$~\cite{y4140a}. In a
subsequent analysis, CDF claimed the observation of the $Y(4140)$
with a statistical significance greater than $5\sigma$ with a mass
of $[4143.4^{+2.9}_{-3.0}({\rm stat}) \pm 0.6({\rm
syst})]$~MeV/$c^{2}$ and a width of $[15.3^{+10.4}_{-0.1}({\rm
stat}) \pm 2.5({\rm syst})]$~MeV~\cite{y4140b}. However,
the existence of the $Y(4140)$ was not confirmed by the Belle~\cite{y4350} or
LHCb~\cite{lhcb} collaborations in the same process, nor by the
Belle collaboration in two-photon production~\cite{y4350}.
Recently, the CMS~\cite{y4140cms} and D0~\cite{d0y4140}
collaborations reported on analyses of $B^{+} \rightarrow
\phijpsi K^{+}$, where an accumulation of events is observed in the
$\phijpsi$ invariant mass distribution, with resonance parameters
consistent with those of the CDF measurement. The BABAR
collaboration also investigated the same decay mode, and found no
evidence for the $Y(4140)$~\cite{babar}.

Being well above the open charm threshold, the narrow structure
$Y(4140)$ is difficult to be interpreted as a conventional
charmonium state~\cite{y4140exp:1}, while it is a good candidate
for a
molecular~\cite{y4140exp:2,y4140exp:3,y4140exp:4,y4140exp:5,y4140exp:6,y4140exp:7},
$c\bar{c}s\bar{s}$ tetraquark~\cite{y4140exp:8}, or charmonium
hybrid state~\cite{y4140exp:3}. A detailed review on the $Y(4140)$
is given in Ref.~\cite{phijpsi_review}. The $Y(4140)$ is the first
charmoniumlike state decaying into two vector mesons consisting of
$c\bar{c}$ and $s\bar{s}$ pairs. Since both the $\phi$ and $\jpsi$
have $J^{PC}=1^{--}$, the $\phijpsi$ system has positive C-parity,
and can be searched for through radiative transitions of $Y(4260)$
or other $1^{--}$ charmonium or charmoniumlike states. The author
of Ref.~\cite{y4140exp:3} found that the partial width of the
radiative transition $Y(4260) \to \gamma Y(4140)$ may be up to
several tens of keV if both the $Y(4260)$ and $Y(4140)$ are hybrid
charmonium states. The data samples collected at center-of-mass
(CM) energies near the $Y(4260)$ at the BESIII experiment can be
used to search for such transitions.

The structure of this paper is as follows. In Sec.~\ref{sec:det},
the setup for the BESIII experiment and details of the data
samples are given. In Sec.~\ref{sec:sel}, event selections for
$\phi J/\psi$ events are described for three different decay modes of
the $\phi$ meson. Section~\ref{sec:cross} details the upper limit
calculations for the production of $Y(4140)$, while
Sec.~\ref{sec:syst} describes the systematic errors of the
measurement. A short summary of the results is given in
Sec.~\ref{sec:summ}.

\section{Data and Monte Carlo samples}
\label{sec:det}

In this paper, we present results of a search for $Y(4140)$ decays
into $\phijpsi$ through the process $\EE \to \gamma \phijpsi$ with
data taken at CM energies of $\sqrt{s} = 4.23$, 4.26, and $4.36$~GeV.
The data samples were collected with the BESIII
detector operating at the BEPCII storage ring~\cite{bepc2}.
The integrated luminosity of these data samples are measured by
using large-angle Bhabha scattering with an uncertainty of $1.0\%$~\cite{pipihc}.
The luminosities of the data samples are 1094, 827, and
$545$~pb$^{-1}$, for $\sqrt{s} = 4.23$, 4.26, and 4.36~GeV,
respectively.

The BESIII detector, described in detail in Ref.~\cite{bepc2}, has
a geometrical acceptance of $93\%$ of $4\pi$. A small-cell
helium-based main drift chamber (MDC) provides a charged particle
momentum resolution of $0.5\%$ at $1$~GeV/$c$ in a $1$ T magnetic
field, and supplies energy loss ($dE/dx$) measurements with a
resolution better than $6\%$ for electrons from Bhabha scattering.
The electromagnetic calorimeter (EMC) measures photon energies
with a resolution of $2.5\%$ ($5\%$) at $1.0$~GeV in the barrel
(endcaps). Particle identification (PID) is provided by a
time-of-flight system (TOF) with a time resolution of $80$~ps
($110$~ps) for the barrel (endcaps). The muon system,
located in the iron flux return yoke of the magnet,
provides
$2$~cm position resolution and detects muon tracks with momentum
greater than $0.5$~GeV$/c$.

The {\sc geant}4-based~\cite{geant4} Monte Carlo (MC) simulation software {\sc
boost}~\cite{boost} includes the geometric description of the
BESIII detector and a simulation of the detector
response.  It is used to optimize
event selection criteria, estimate backgrounds and evaluate the
detection efficiency.  For each energy point, we generate signal MC samples of $\EE
\rightarrow \gamma Y(4140)$, $Y(4140) \rightarrow \phijpsi$ uniformly
in phase space, where the $\phi$ decays to $\kk/\KK/\pipipi$ and
the $\jpsi$ decays to $\EE/\MM$.
The decays of $\phi \to \kk$ and $\KK$ are modeled as a vector
particle decaying to two pseudoscalars ({\sc evtgen}~\cite{evtgen} model {\sc vss}),
 and the decay $\phi\to \rho \pi$ is modeled as a vector particle decaying to a vector and a scalar ({\sc
vvs\_pwave} model), and all the other processes are generated
uniformly in phase space. Effects of initial state
radiation (ISR) are simulated with {\sc kkmc}~\cite{kkmc}, where
the Born cross section of $\EE \rightarrow \gamma Y(4140) $ is
assumed to follow the $Y(4260) \rightarrow \pp \jpsi$
line shape~\cite{pdg}. Final state radiation (FSR) effects
associated with charged particles are handled with {\sc
photos}~\cite{photos}.

To study possible background contributions, MC samples of inclusive
$Y(4260)$ decays, equivalent to the integrated luminosity of data, are
also generated at $\sqrt{s}=4.23$, $4.26$ and $4.36$ GeV.  In these
simulations the $Y(4260)$ is allowed to decay generically, with the
main known decay channels being generated using
\textsc{evtgen} with branching fractions set to world
average values~\cite{pdg}.  The remaining events associated with
charmonium decays are generated with {\sc lundcharm}~\cite{lundcharm}
while continuum hadronic events are generated with {\sc
  pythia}~\cite{pythia}. QED events such as Bhabha, dimuon and digamma
are generated with {\sc kkmc}~\cite{kkmc}.

\section{Event selection}
\label{sec:sel}

For each charged particle track, the polar angle in the MDC
must satisfy  $|\cos\theta|<0.93$, and the point of closest
approach to the $\EE$ interaction point (IP) must be within
$\pm$10~cm in the beam direction and within $\pm$1~cm in the plane
perpendicular to the beam direction, except for the $\pi^+ \pi^-$
pair from $\ks$ decays. Since leptons from the $\jpsi$
decays are kinematically well separated from other charged tracks,
tracks with momenta larger than $1.0$~GeV$/c$ in the
laboratory frame are assumed to be leptons. We use the energy
deposited in the EMC to separate electrons from muons. For muon
candidates, the deposited energy is less than 0.4~GeV, while for
electrons it is larger than $1.0$~GeV. EMC showers identified as
photon candidates must satisfy the following
requirements. The minimum required energy deposited in the EMC is
$25$~MeV for the barrel ($|\cos\theta|<0.8$) and $50$~MeV for the
endcaps ($0.86<|\cos\theta|<0.92$). To eliminate showers
associated with
charged particles, e.g. from bremsstrahlung, a photon must be
separated by at least 20 degrees from any charged track. The
timing information from the EMC is also required to be in 0-700 ns
to suppress electronic noise and energy deposits unrelated to
signal events.

\subsection{$\phi \to \kk$}

For the $\phi \to \kk$ decay mode, the momenta of the kaons are about
$0.2$~GeV/$c$ in the laboratory frame. The detection efficiency
for low momentum kaons is very small. In order to increase the
efficiency, only one kaon is required to be found and to pass through
the PID selection using both $dE/dx$ and TOF information. To improve the
mass resolution and suppress backgrounds, a one-constraint (1C)
kinematic fit is performed with the $\gamma K^{+} K^{-} \LL$ ($\ell=e$
or $\mu$) hypothesis, constraining the missing mass to the Kaon mass, and the
$\chi^{2}$ is required to be less than $25$. This value is
determined by maximizing the figure of merit (FOM)~$S/\sqrt{S+B}$,
where $S$ refers to the number of signal events from the signal MC
simulation and $B$ is the number of background events from the
inclusive MC sample. For the signal cross section, we use the
upper limit determined in this analysis as input. 
The $\chi^{2}$ requirement depends weakly on the cross section of
signal. If there are two
kaons or more than one good photon candidate, the combination with
the smallest $\chi^{2}$ is retained.

\begin{figure*}[htbp]
\centering
\begin{overpic}[width=0.49\textwidth]{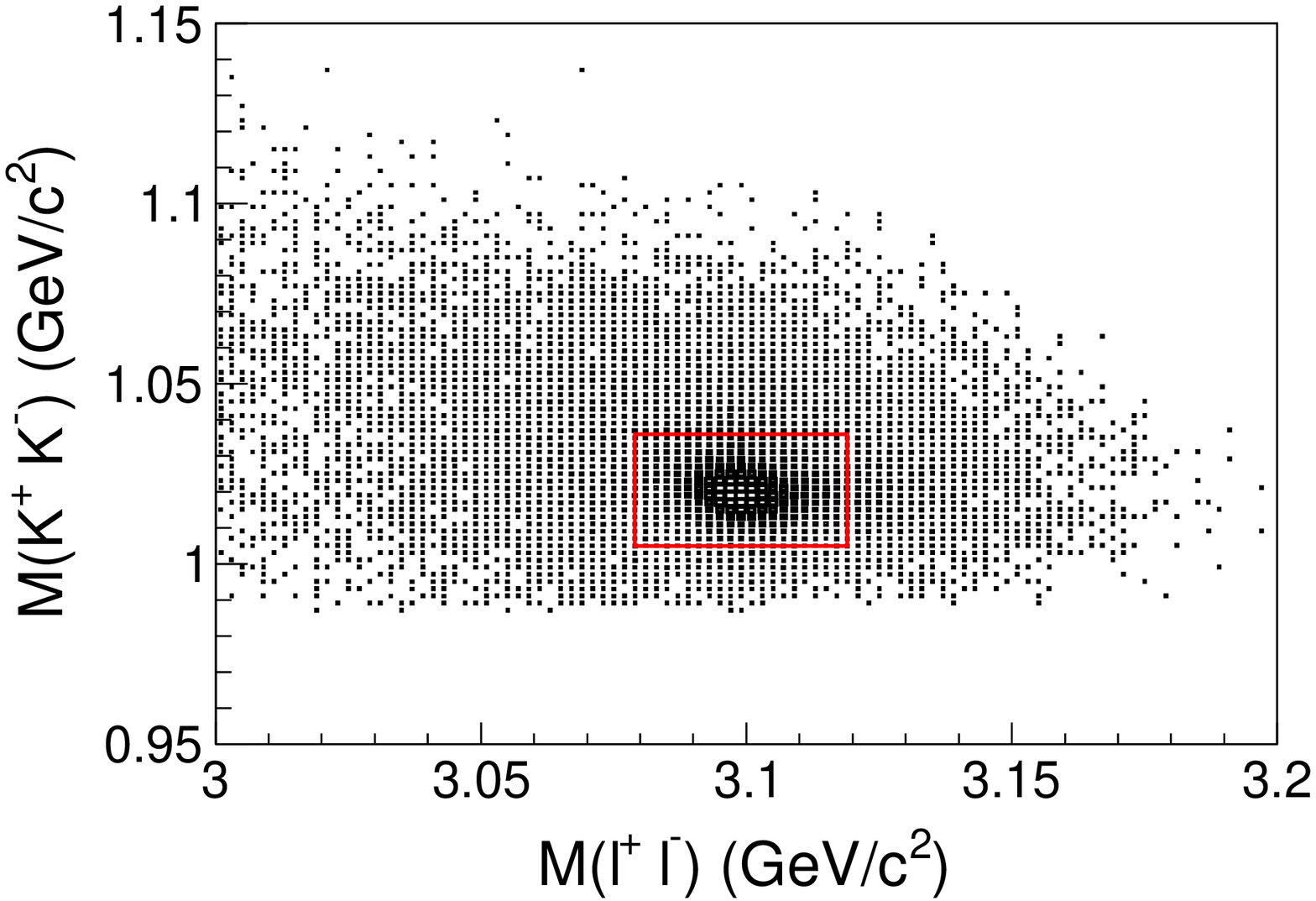}
\put(25,55){(a)}
\end{overpic}
\begin{overpic}[width=0.49\textwidth]{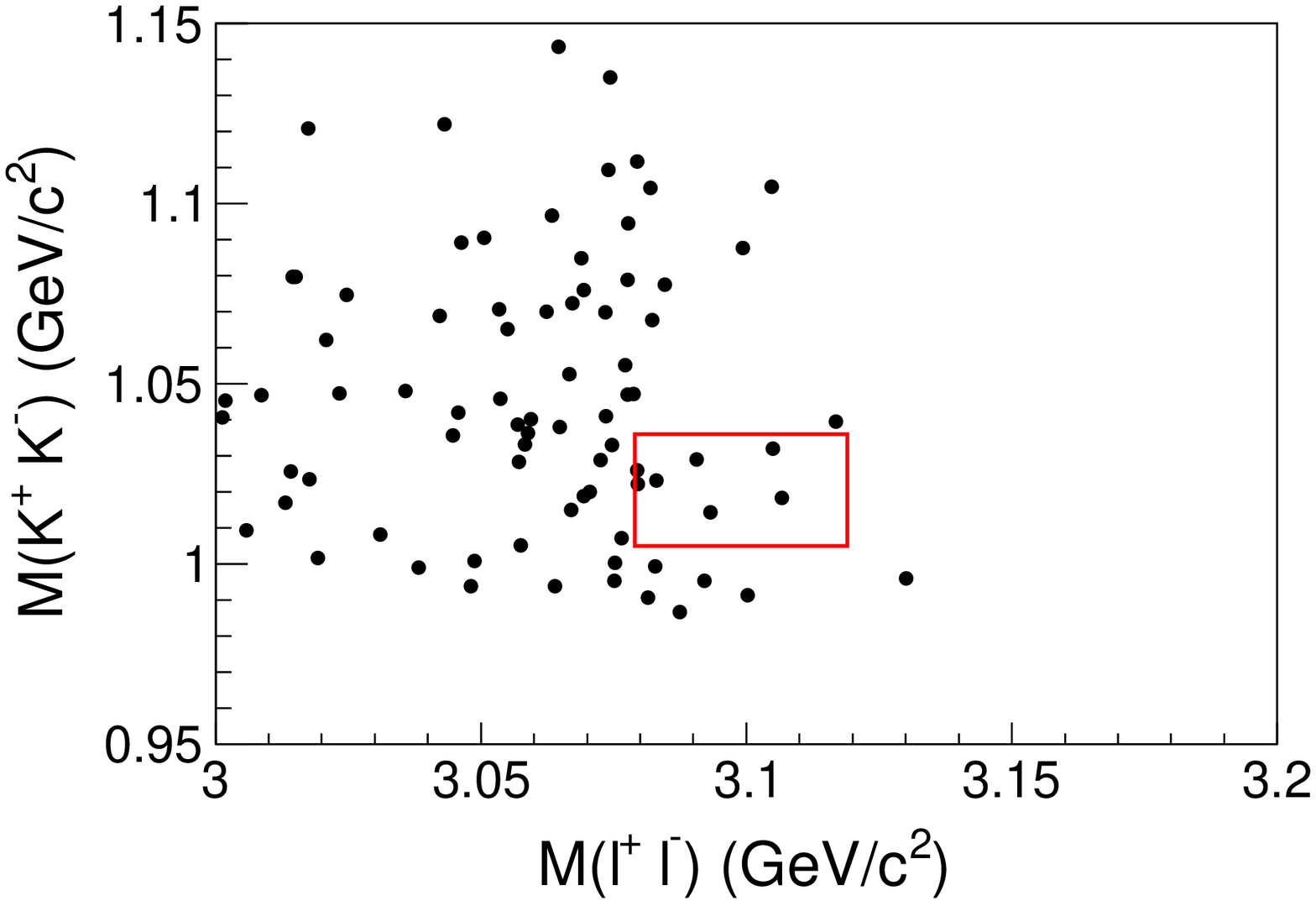}
\put(25,55){(b)}
\end{overpic}\\
\begin{overpic}[width=0.49\textwidth]{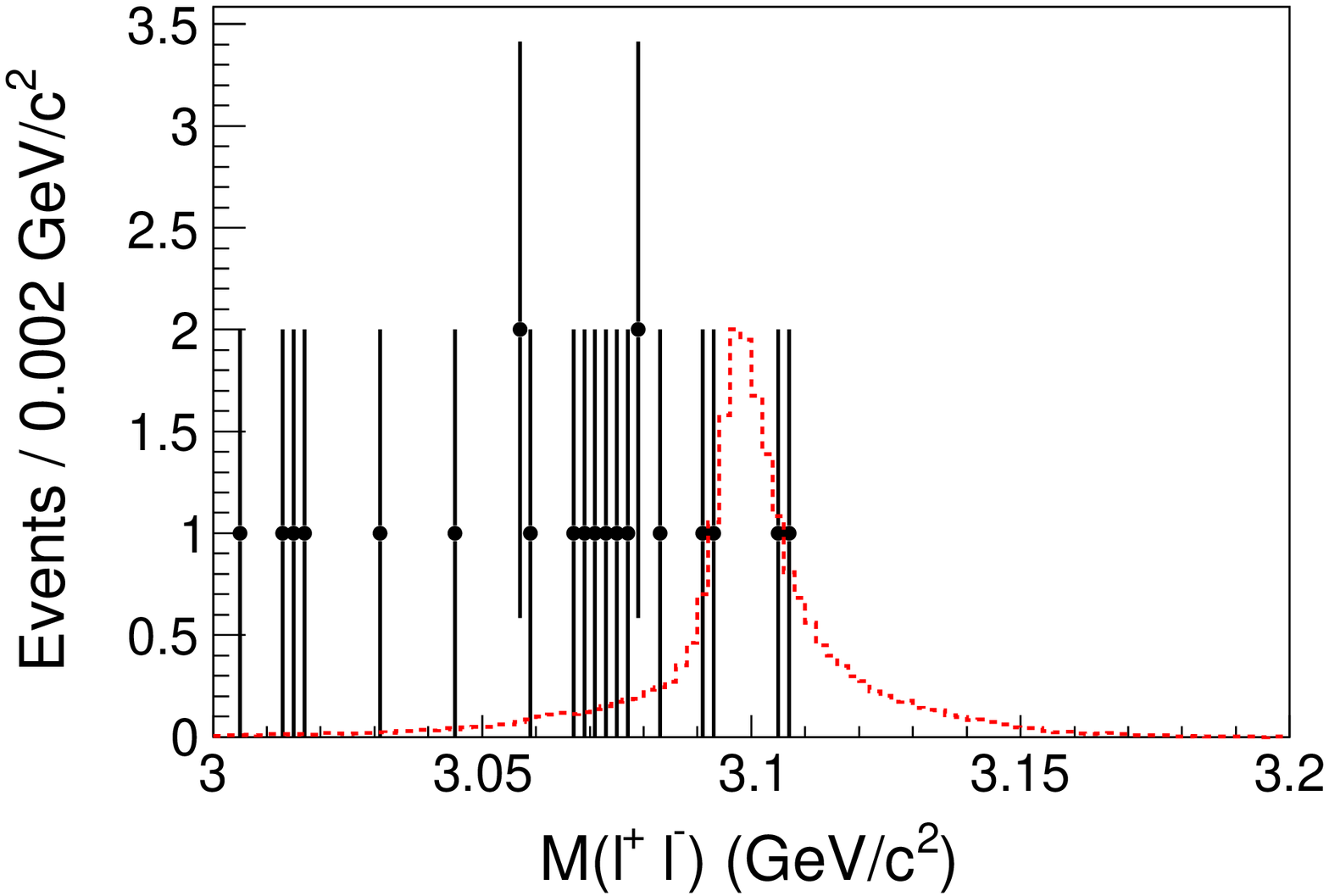}
\put(25,55){(c)}
\end{overpic}
\begin{overpic}[width=0.49\textwidth]{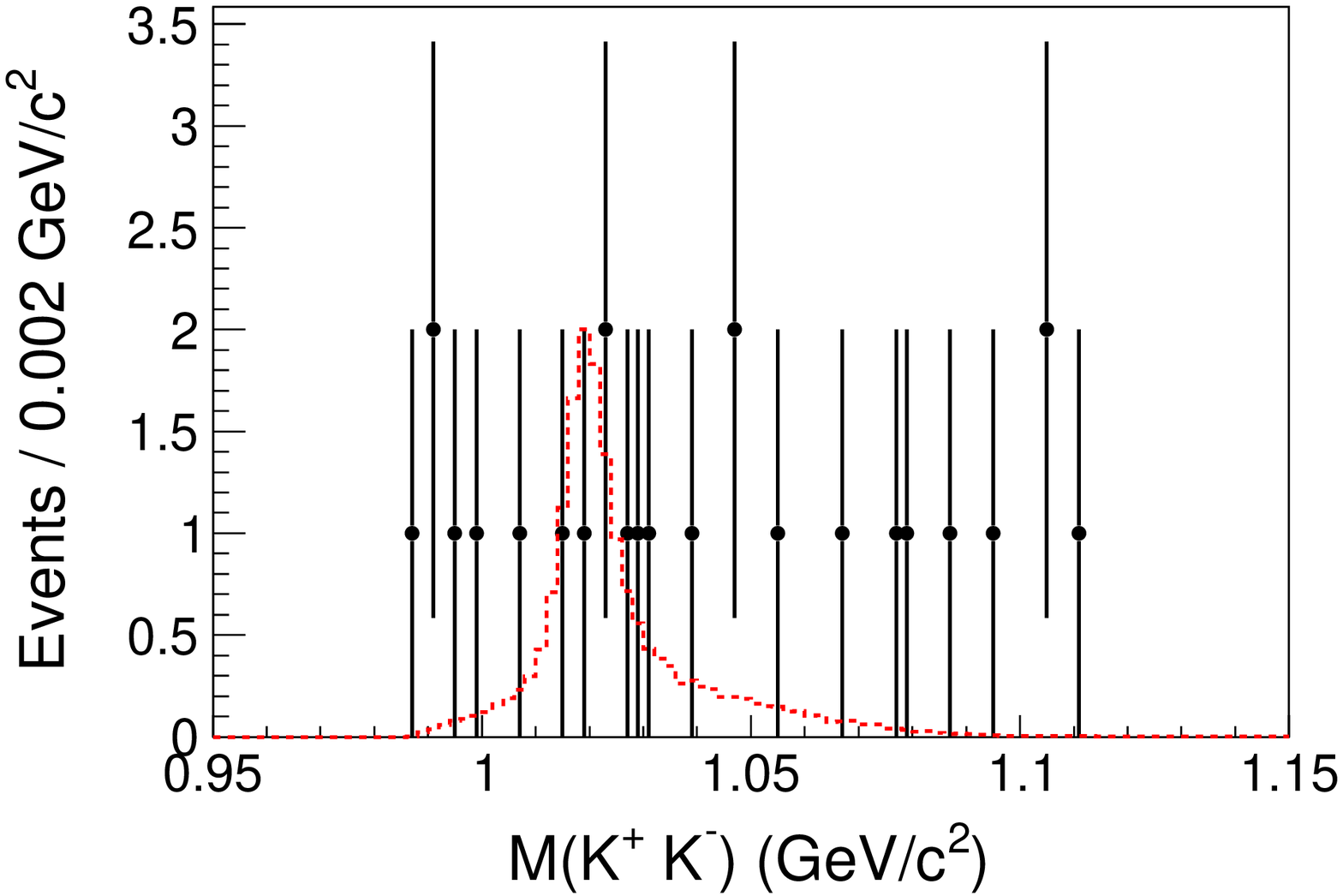}
\put(25,55){(d)}
\end{overpic}
\caption{Scatter plots for (a) signal MC, (b) data at 4.26 GeV and (c) the projections along $M(\LL)$ in $\phi$ mass window and (d) the projections along $M(\kk)$ in $\jpsi$ mass window. Red box shows mass windows of $\phi$ and $\jpsi$. Red dashed histogram shows the MC simulated shape (not normalized).}
\label{fig_scatter}
\end{figure*}

After imposing the requirements above, we use mass windows around the
$\jpsi$ and $\phi$ to select signal events. The mass windows are
defined as $[\mu-W, \mu+W]$, where $\mu$ and $W$ are the mean value
and full width at half maximum (FWHM) of the invariant mass
distributions of signal events from the MC simulation. The values of
$\mu$ and $W$ for each of the different decay modes of the $\phi$
meson considered in this analysis are listed in
Table~\ref{tab:fit_mc}. Figure~\ref{fig_scatter} shows the scatter
plots of $M(\kk)$ vs.\ $M(\LL)$ for MC and data at 4.26 GeV and the
1-D projections. No significant $\gamma \phi \jpsi$ signal is
observed. The dominant background events are $\EE \to \kk\jpsi$ with a
random photon candidate from beam related background cluster, so the mass of
$\jpsi$ is shifted by about 30 MeV/$c^{2}$ to the lower side. About 0.4\%
of these events will leak into the $\jpsi$ mass window,
but in the $M(\phijpsi)$ distribution, they accumulate at about 30
MeV/$c^{2}$ below the CM energy, far away from the nominal mass
of the $Y(4140)$.

\begin{table}[htbp]
\caption{The mean ($\mu$) and FWHM ($W$) of the $\jpsi$ and $\phi$
mass distributions, and the mass windows of the $\jpsi$ and $\phi$
signals. All values are in units of MeV/$c^{2}$.}
\label{tab:fit_mc}
\begin{tabular}{c c c c}
  \hline \hline
  mode & $\mu(\jpsi$) & $W(\jpsi)$ & Mass window  \\
  \hline
  $\phi \rightarrow \kk$ & $3098.9 \pm 0.1$ & $19.8 \pm 0.1$ & 3079-3119    \\
  $\phi \rightarrow \KK$ & $3099.1 \pm 0.1$ & $20.5 \pm 0.1$ & 3078-3120    \\
  $\phi \rightarrow \pipipi$ & $3101.1\pm 0.1$ & $18.6 \pm 0.1$ & 3082-3120 \\
  \hline
    mode &  $\mu(\phi$) & $W(\phi)$ & Mass window \\
  \hline
  $\phi \rightarrow \kk$ & $1020.1 \pm 0.1$ & $15.1 \pm 0.1$ & 1005-1036 \\
  $\phi \rightarrow \KK$ & $1019.8 \pm 0.1$ & $13.9 \pm 0.1$ & 1005-1034 \\
  $\phi \rightarrow \pipipi$ & $1019.1 \pm 0.1$ & $16.8 \pm 0.1$ & 1002-1036 \\
  \hline \hline
\end{tabular}
\end{table}

\begin{figure*}[htbp]
\centering
\begin{overpic}[width=0.49\textwidth]{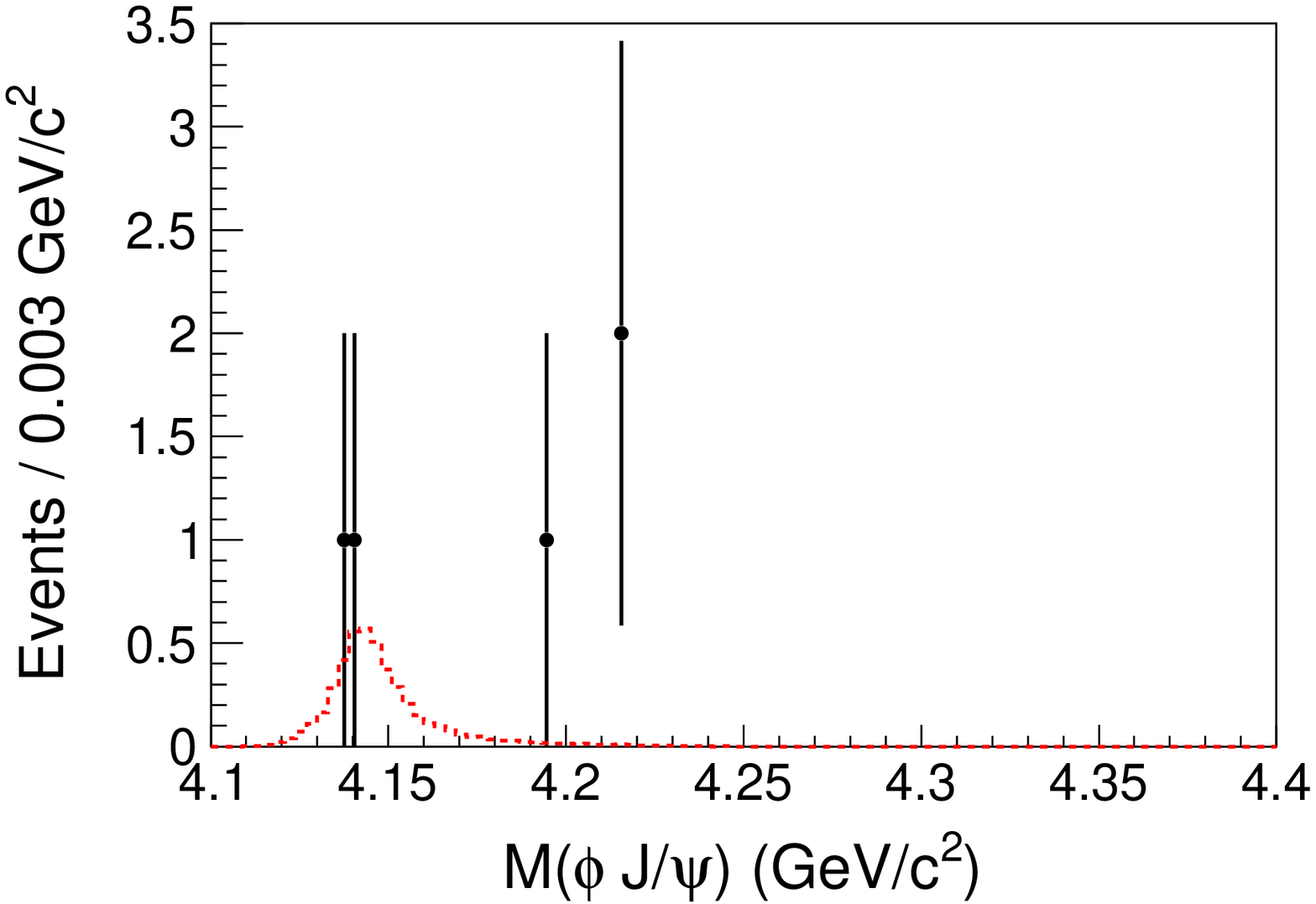}
\put(25,55){(a)}
\end{overpic}
\begin{overpic}[width=0.49\textwidth]{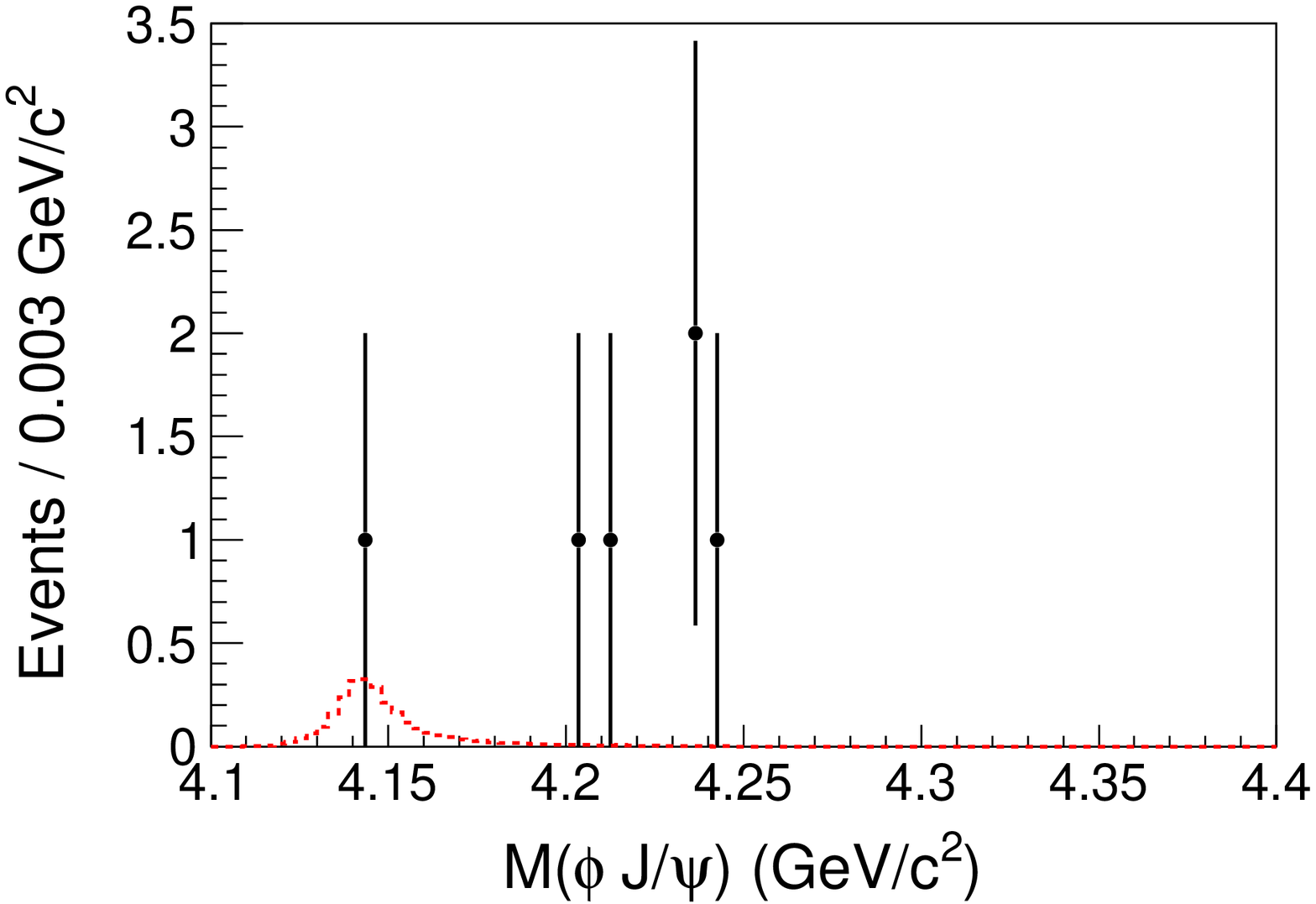}
\put(25,55){(b)}
\end{overpic}\\
\begin{overpic}[width=0.49\textwidth]{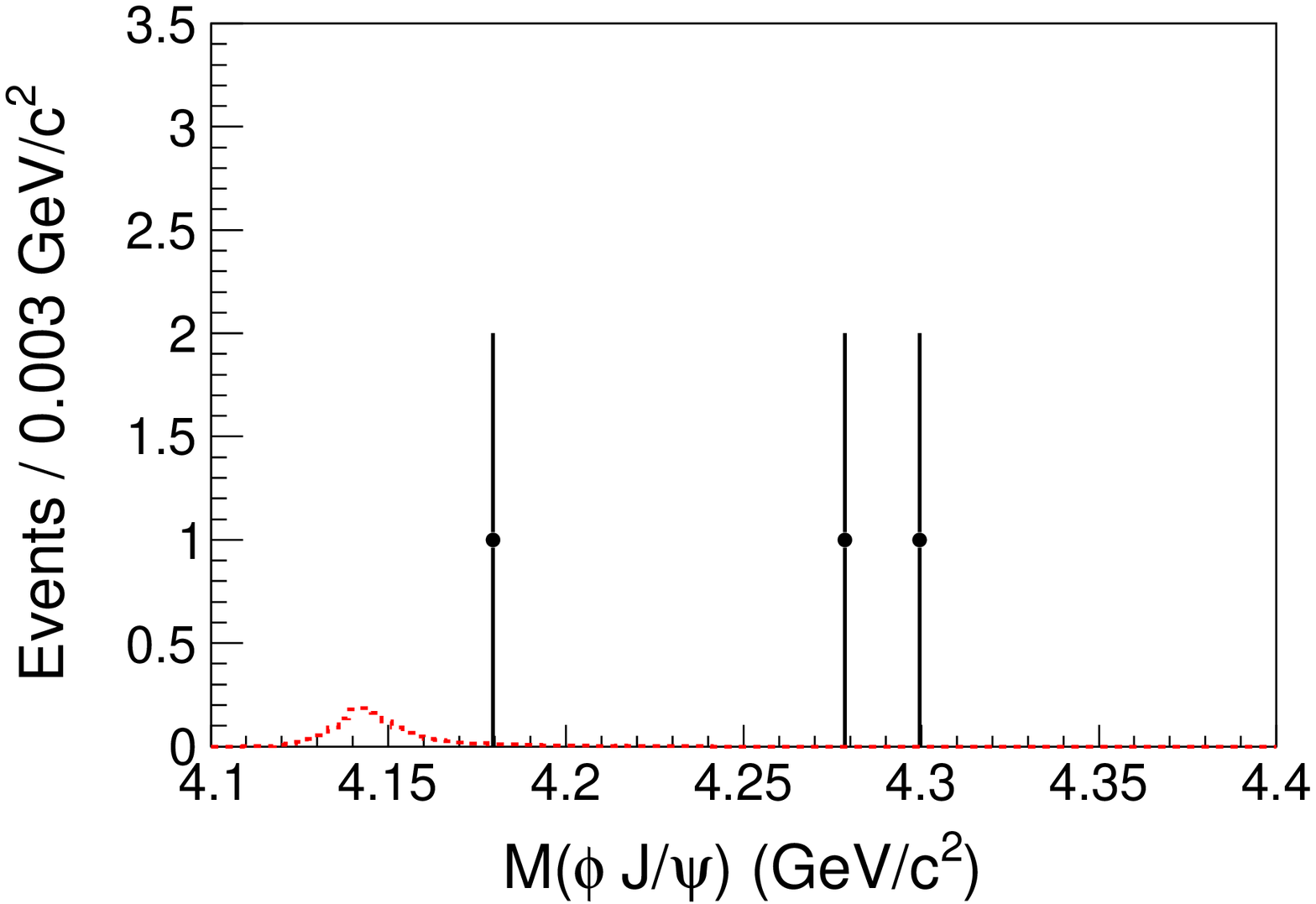}
\put(25,55){(c)}
\end{overpic}
\begin{overpic}[width=0.49\textwidth]{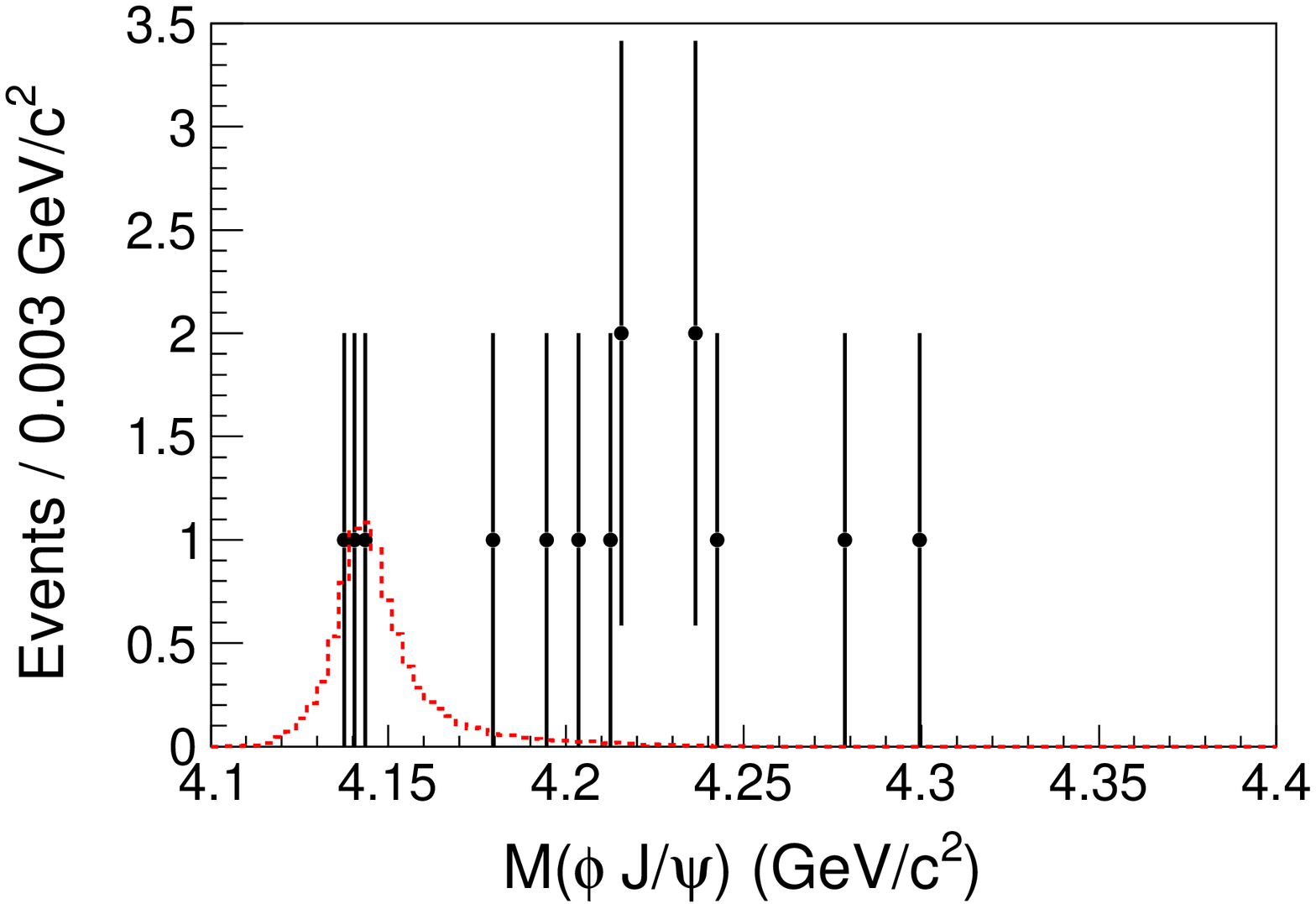}
\put(25,55){(d)}
\end{overpic}
\caption{Distribution of $M(\phi\jpsi)$ with $\phi$ decays to
$\kk$ from data collected at (a) $4.23$, (b) $4.26$, (c) $4.36$~GeV
 and (d) the sum of three data samples. The red dashed histograms
represent signal MC samples scaled to the measured upper limits. }
\label{fig:kkjpsi_phijpsi}
\end{figure*}

The invariant mass distributions of the $\phijpsi$ candidates after
all event selection criteria have been applied are shown in
Fig.~\ref{fig:kkjpsi_phijpsi}, for the three data samples and the sum
of them.  Here we use $M(\phi\jpsi)=M(K^{+} K^{-} \ell^+
\ell^-)-M(\LL)+m_{\jpsi}$ to partially cancel the mass resolution
of the lepton pair, where $m_{\jpsi}$ is the nominal mass
of the $\jpsi$~\cite{pdg}.

There are no events left from the inclusive MC sample after
applying all of the above selections. Since there are two high
momentum leptons in the final state and the BESIII PID can
separate the low momentum kaon from other particles very well, the
possible backgrounds must have a $\kk$ pair and two high-momentum
charged tracks. Exclusive MC samples of the processes $\EE \to \kk \jpsi, \kk
\pipi, \kk \pipipi$ and $\phi \pipi$ are generated and analyzed
with more than $100,000$ events each (corresponding to a cross section
of 200 pb), and we confirm that no events
are selected as the $Y(4140)$ signal. The cross sections
of these final states have been measured to be of a few or a few tens of pb level~\cite{belle:kkj,babar:kkpp,babar:kkppp,cleo:kkj} in the energy range of interest. Backgrounds due to one photon from $\pi^{0}$ or $\eta$
decays being misidentified as the radiative photon were checked for in the inclusive MC sample
and found to be negligible.

Three-body process $\EE \to \gamma \phi \jpsi$ and four-body process
$\gamma \kk \jpsi$ are studied with MC simulation.  Even though the
cross sections of these non-resonant channels are expected to be
small, we cannot rule out the possibility that the three events
observed in the $Y(4140)$ signal region (as shown in
Fig.~\ref{fig:kkjpsi_phijpsi}) are from non-resonant processes.

\subsection{$\phi \to \KK$}

For the $\phi \to \KK$ mode, the $\ks$ is reconstructed with its
decay to $\pipi$. The pions from the decay of $\ks$ can also be
kinematically well separated from the leptons, and charged tracks with
momenta less than 0.6 GeV/$c$ in the laboratory frame are assumed
to be pions. Since the $K^0_S$ has a relatively long lifetime, it
travels a measurable distance before it decays. We perform a
secondary vertex fit on the two charged pions to improve the mass
resolution, but no extra $\chi^2$ requirement is applied. The
fitted mass and FWHM of the $\pp$ invariant mass spectrum is
determined from the simulation to be $\mu = (497.6\pm0.1)$~MeV/$c^{2}$
and $W = (3.3\pm0.1)$~MeV/$c^{2}$, respectively,
and we select candidates in the mass range
$[\mu-W, \mu+W]$.  Since the $K_{L}^{0}$ is difficult to be
detected at BESIII, we only require that there are two pions and two
leptons in the final state. Then the event is kinematically fitted
to the hypothesis $\gamma \KK\LL$, with the missing mass constrained to the
nominal $K_{L}^{0}$ mass~\cite{pdg}. If there is more than one
good photon candidate, the combination with the smallest
$\chi^{2}$ is used, and the $\chi^{2}$ is required to be less than
$20$.

\begin{figure*}[htbp]
\centering
\begin{overpic}[width=0.49\textwidth]{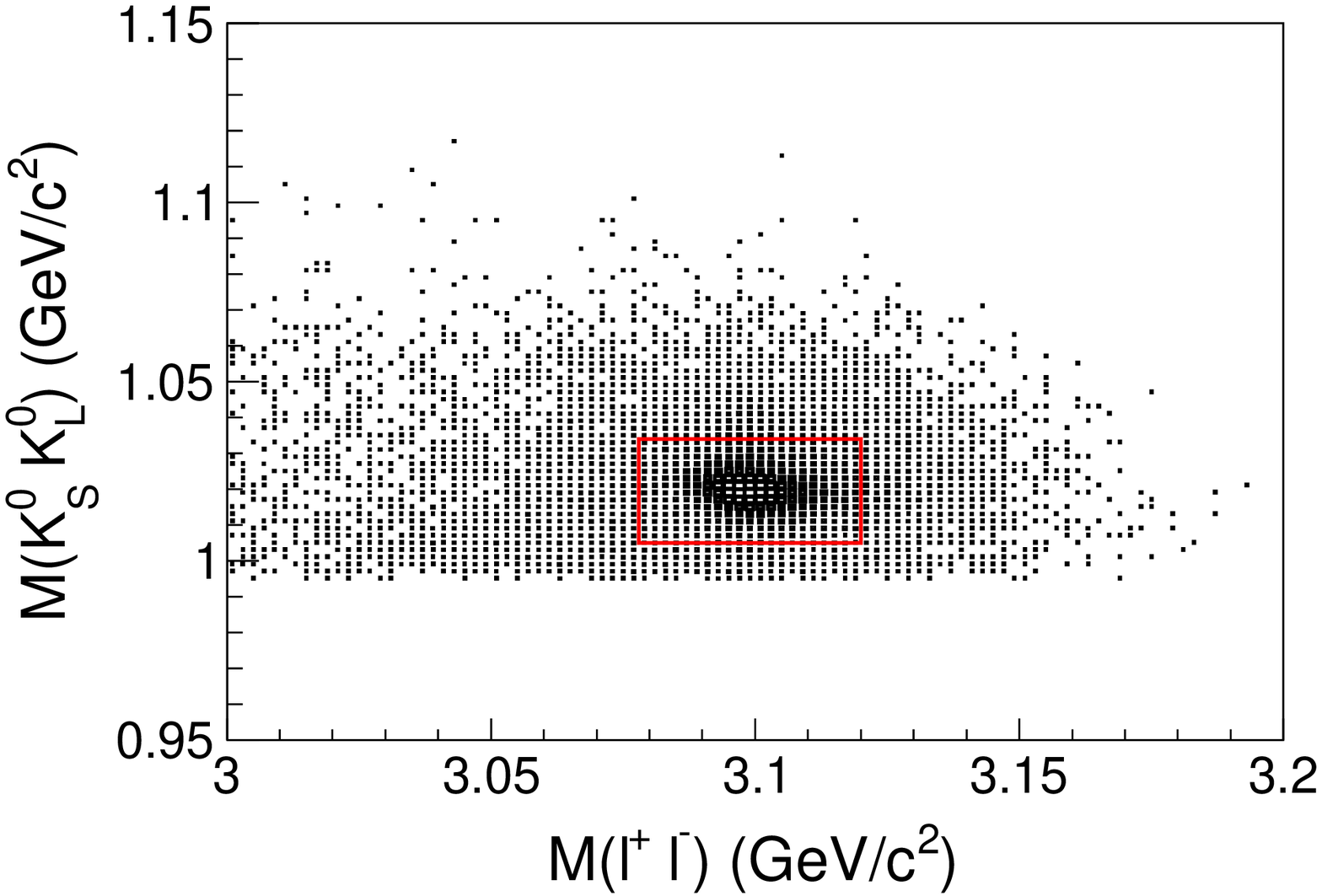}
\put(25,55){(a)}
\end{overpic}
\begin{overpic}[width=0.49\textwidth]{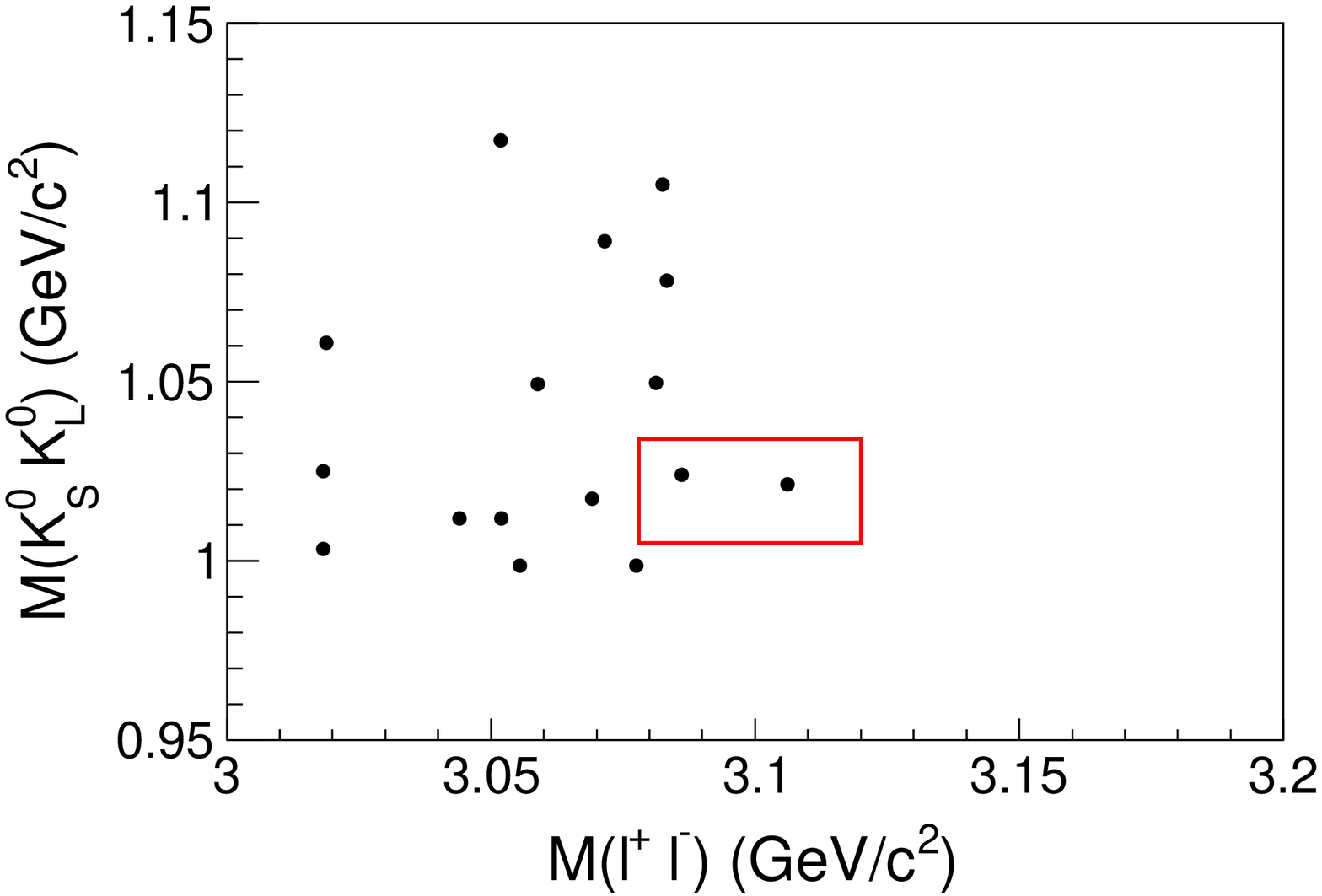}
\put(25,55){(b)}
\end{overpic}\\
\begin{overpic}[width=0.49\textwidth]{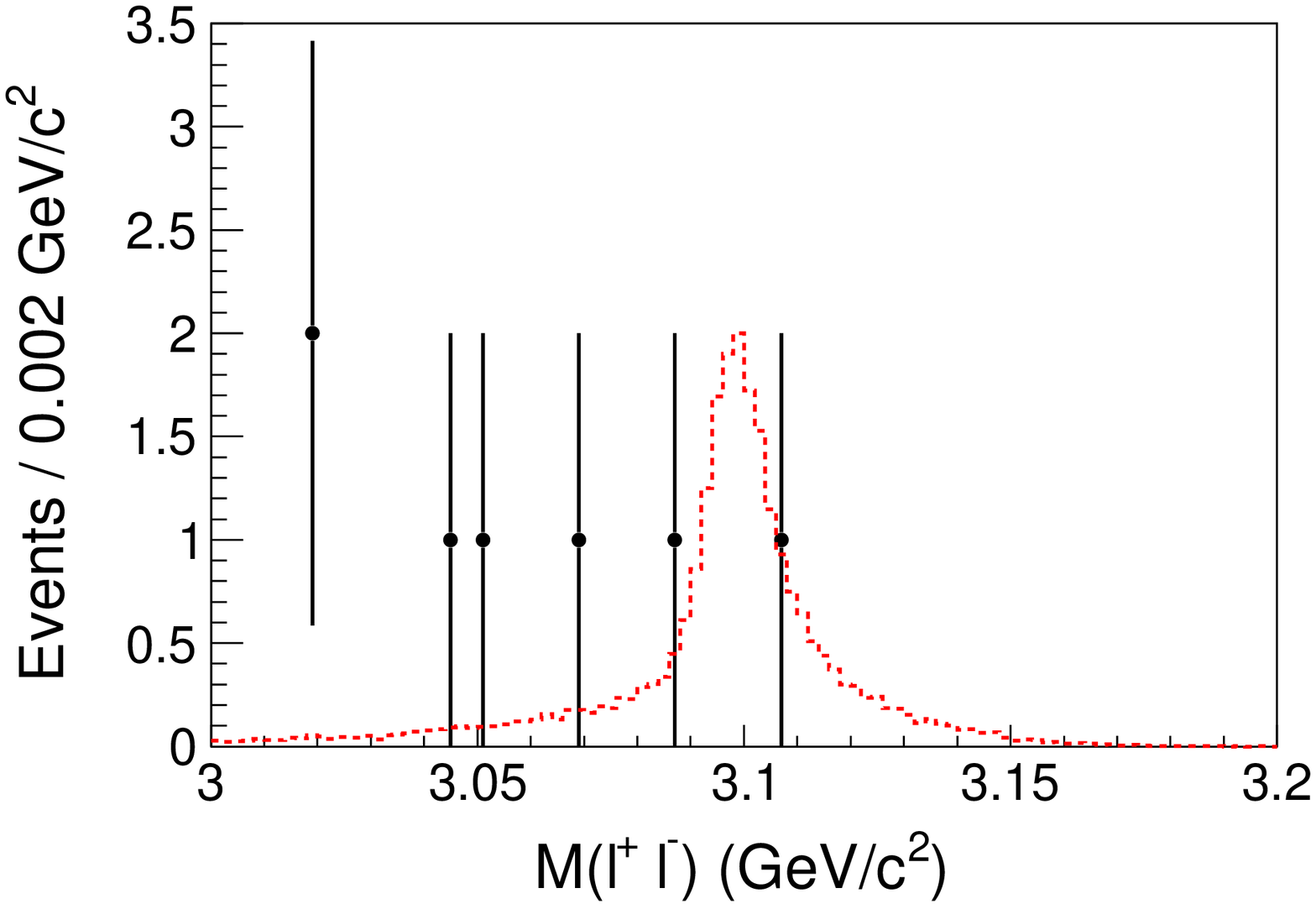}
\put(25,55){(c)}
\end{overpic}
\begin{overpic}[width=0.49\textwidth]{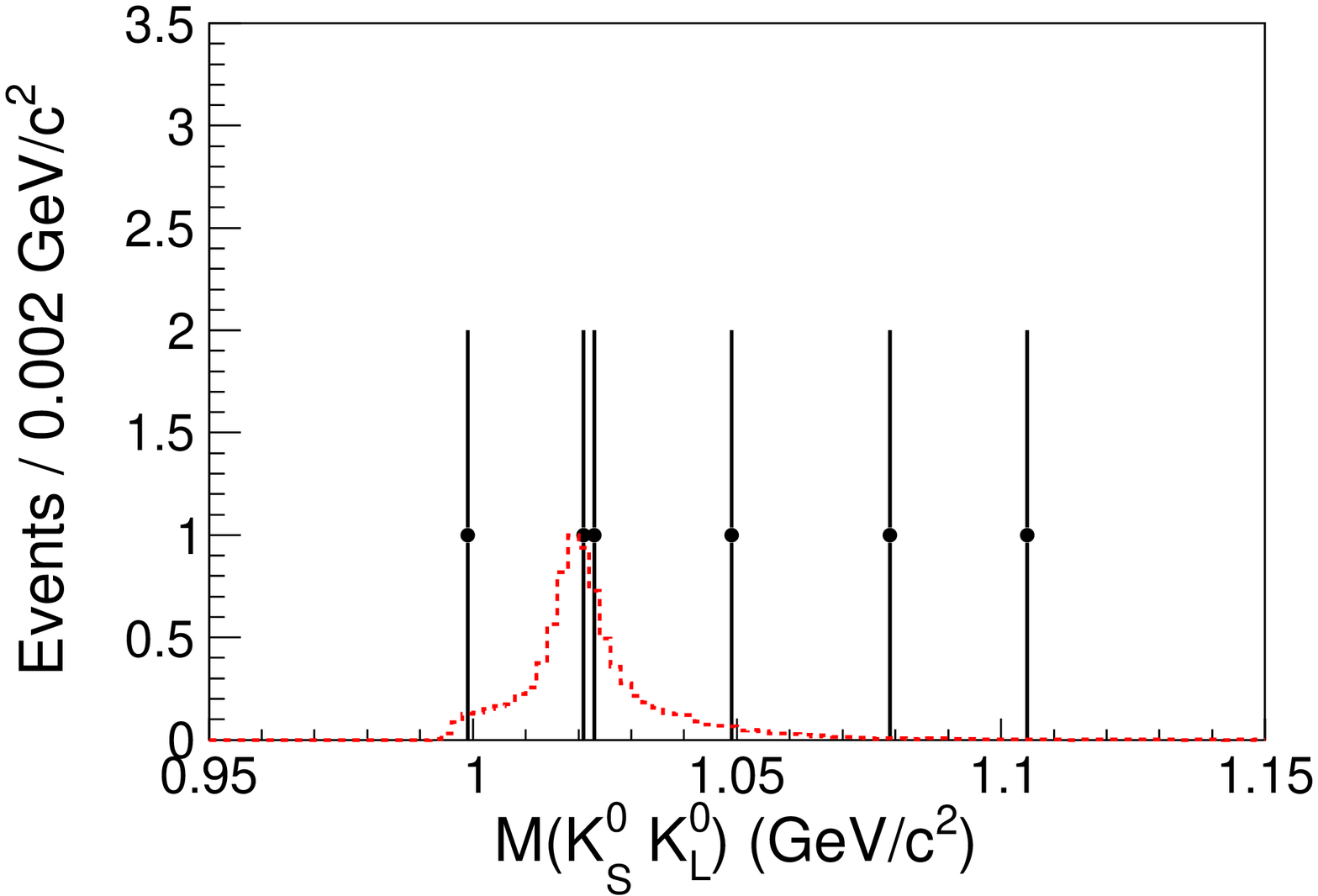}
\put(25,55){(d)}
\end{overpic}
\caption{Scatter plots for (a) signal MC, (b) data at 4.26 GeV and (c)
  the projections along $M(\LL)$ in the $\phi$ mass window, and (d)
  the projections along $M(\KK)$ in the $\jpsi$ mass window. The red
  box shows the mass regions used for $\phi$ and $\jpsi$. The red
  dashed histograms show the MC simulated shape (with arbitrary normalization).}
\label{fig_scatter_kskl}
\end{figure*}

The mass windows around the $\jpsi$ and $\phi$ used to select
signal events are given in
Table~\ref{tab:fit_mc}. Figure~\ref{fig_scatter_kskl} shows the
scatter plots of $M(\KK)$ vs.\
$M(\LL)$ for MC and data at 4.26 GeV and the 1-D projections. The
dominant background events are from $\EE \to
\KK\jpsi$ with a random photon candidate, so the mass
of $\jpsi$ is shifted too, as in the $\phi \to \kk$ mode.

\begin{figure*}[htbp]
\begin{center}
\begin{overpic}[width=0.49\textwidth]{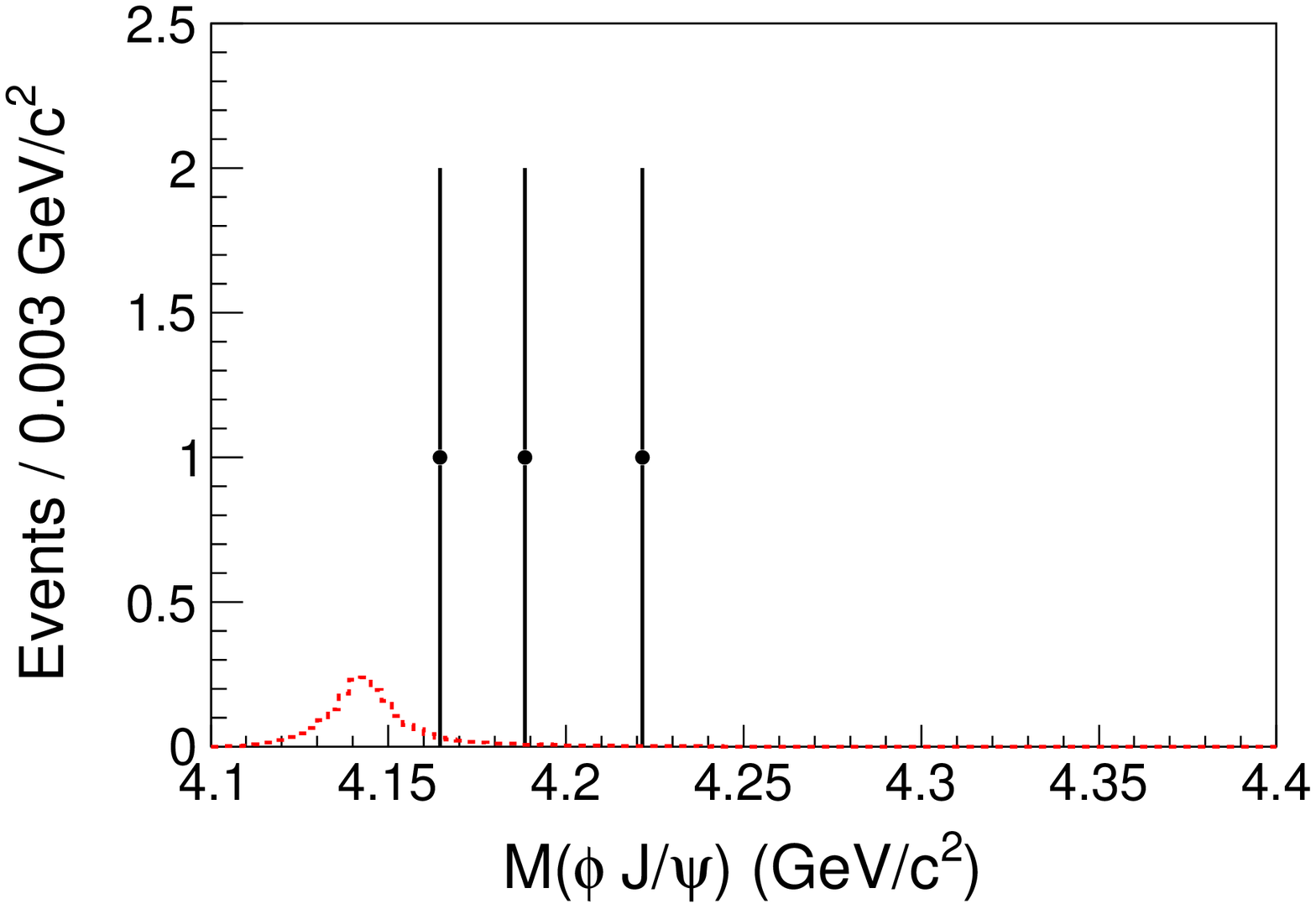}
\put(25,55){(a)}
\end{overpic}
\begin{overpic}[width=0.49\textwidth]{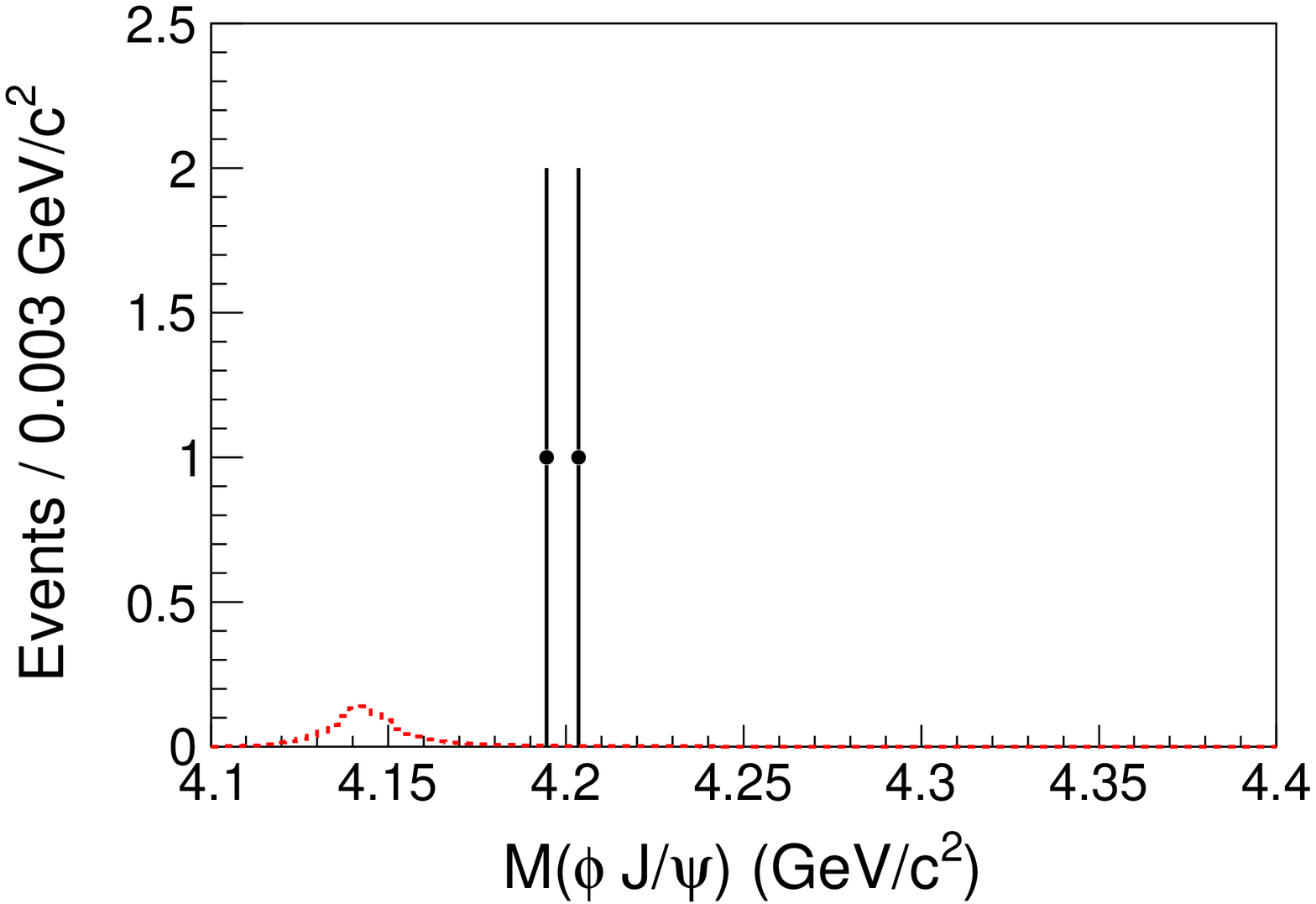}
\put(25,55){(b)}
\end{overpic}\\
\begin{overpic}[width=0.49\textwidth]{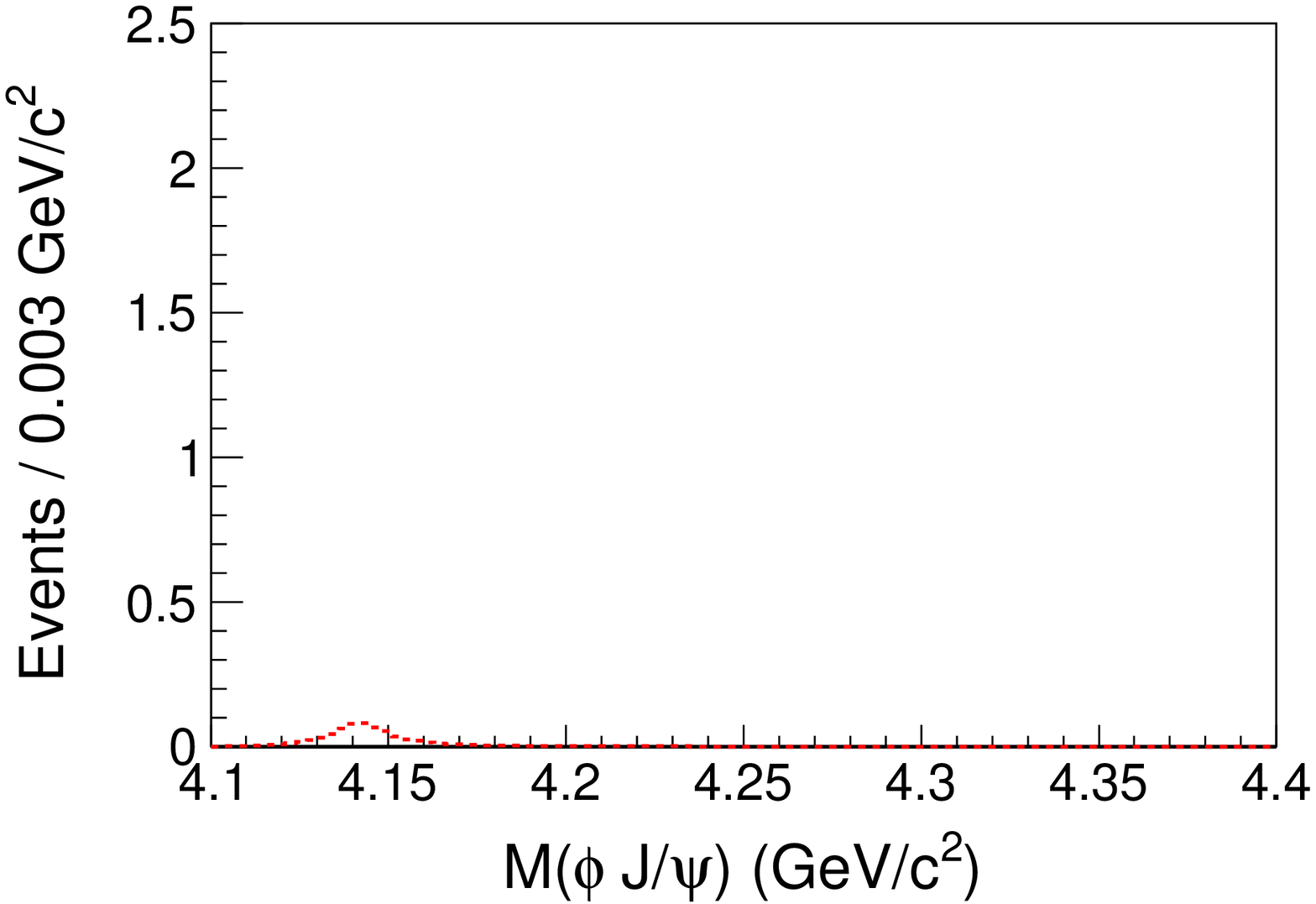}
\put(25,55){(c)}
\end{overpic}
\begin{overpic}[width=0.49\textwidth]{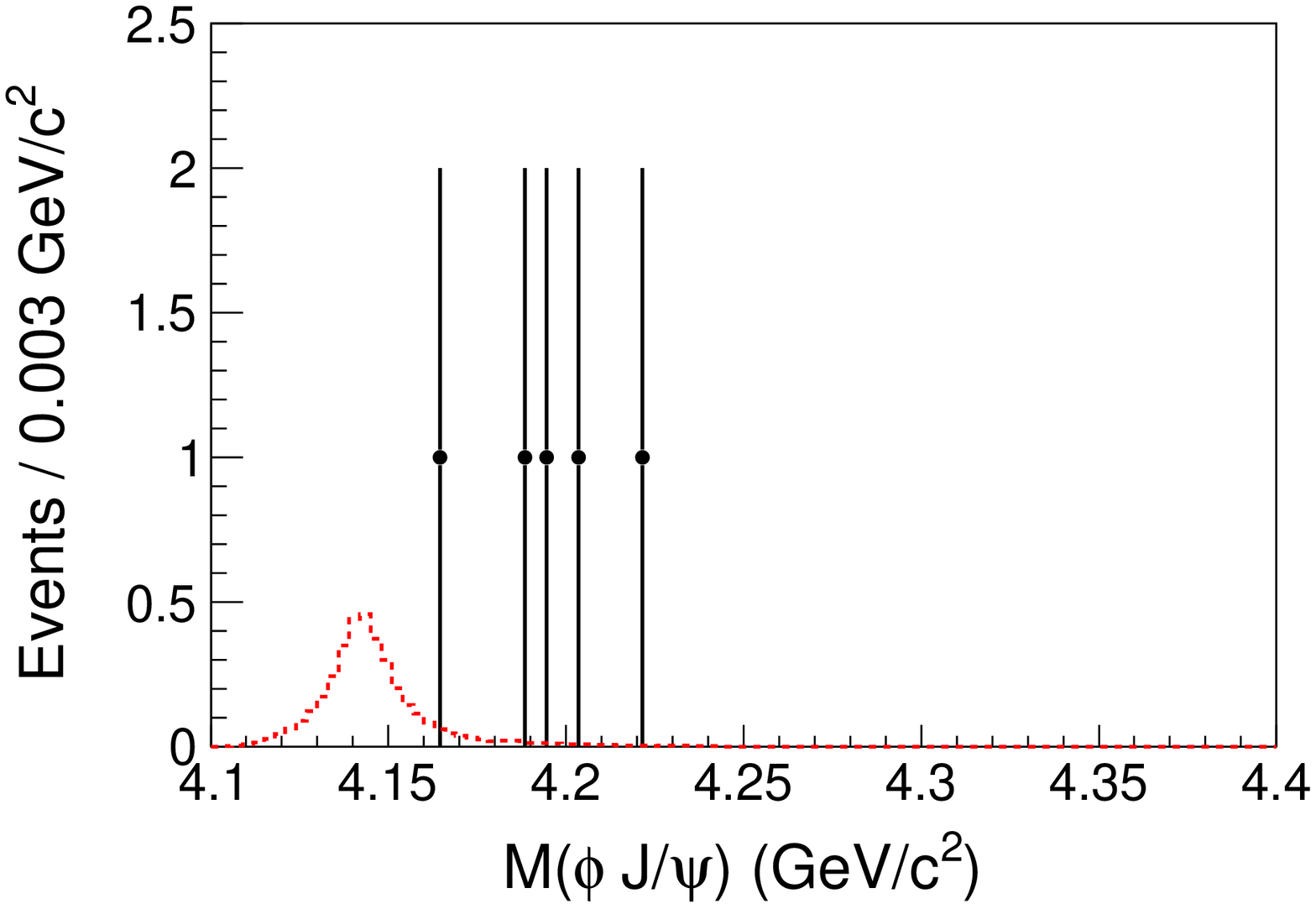}
\put(25,55){(d)}
\end{overpic}
\end{center}
\caption{Distribution of $M(\phi\jpsi)$ with $\phi$ decays to
$\KK$ from data collected at (a) $4.23$, (b) $4.26$, (c) $4.36$~GeV,
and (d) the sum of the three data samples. The red dashed
histograms represent signal MC samples which have been scaled to
the measured upper limits.} \label{fig:phijpsi_kskljpsi}
\end{figure*}

To study possible backgrounds,
 we use the
inclusive MC sample, as well as exclusive MC samples of $\EE \to
\KK \jpsi, \eta \eta \jpsi, \eta \jpsi$ and $\phi \pp$.  No events survive in the
$Y(4140)$ signal region. The size of each exclusive MC samples corresponds
to a production cross section of 200 pb, which is larger than at least a factor of 4 of
the experimental measurements~\cite{belle:kkj,babar:kkpp,belle:etaj,cleo:kkj}. Figure~\ref{fig:phijpsi_kskljpsi} shows the
distribution of $M(\phi\jpsi)=M(\KK
\ell^+\ell^-)-M(\LL)+m_{\jpsi}$ after all the event selection criteria
have been applied, with no obvious $Y(4140)$ or other signals. There
are only 5 events in the sum of three data samples, and none of them is
near the mass of the $Y(4140)$.

\subsection{$\phi \to \pipipi$}

For the $\phi \to \pipipi$ decay mode, the charged pions from the
$\phi$ decays have lower momenta than the leptons from the $\jpsi$
decay, so all charged tracks with momentum less than $0.6$~GeV$/c$
are taken to be pions. We require that there are at least three
good photons in the EMC, and loop over all the combinations to
select three photons with the smallest $\chi^{2}$ of a
four-constraint (4C) kinematic fit, which constrains the
four-momenta of all particles in the final state to be that of the
initial $e^{+} e^{-}$ system.  The $\chi^{2}$ is required to be
less than $40$. We use two photons out of the three to reconstruct
a $\pi^{0}$ candidate, whose invariant mass is nearest to the nominal mass
of the $\pi^{0}$~\cite{pdg}. The fitted mass and FWHM of the
$\pi^{0}$ of signal events from MC simulation are $\mu = (134.1 \pm
0.1)$~MeV/$c^{2}$ and $W = (8.2 \pm 0.1)$~MeV/$c^{2}$, respectively.
We select $\pi^0$ candidates in the mass range $[\mu-W, \mu+W]$, and
the mass windows of $\jpsi$ and $\phi$ from this mode are also
shown in Table~\ref{tab:fit_mc}.

\begin{figure*}[htbp]
\centering
\begin{overpic}[width=0.49\textwidth]{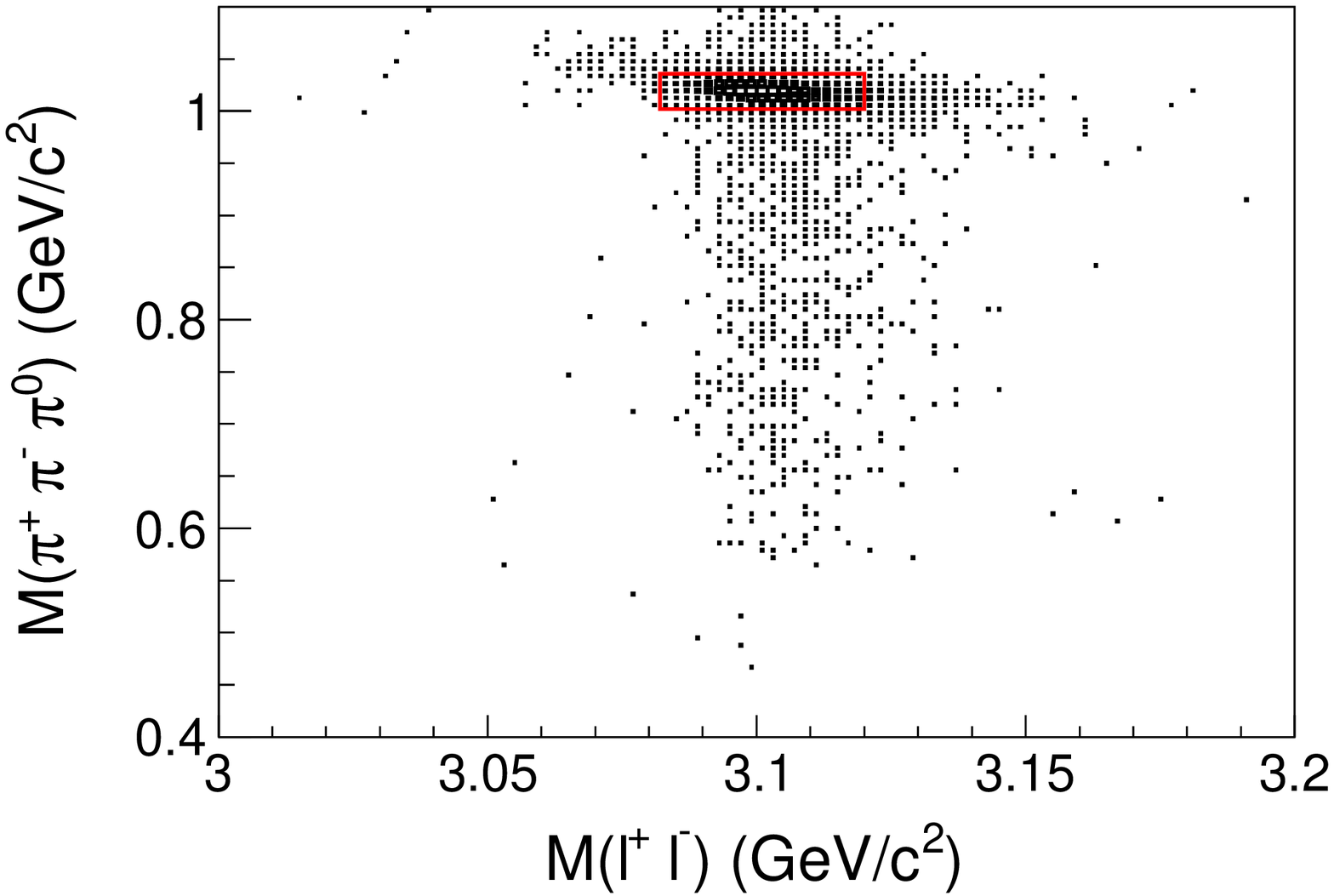}
\put(25,55){(a)}
\end{overpic}
\begin{overpic}[width=0.49\textwidth]{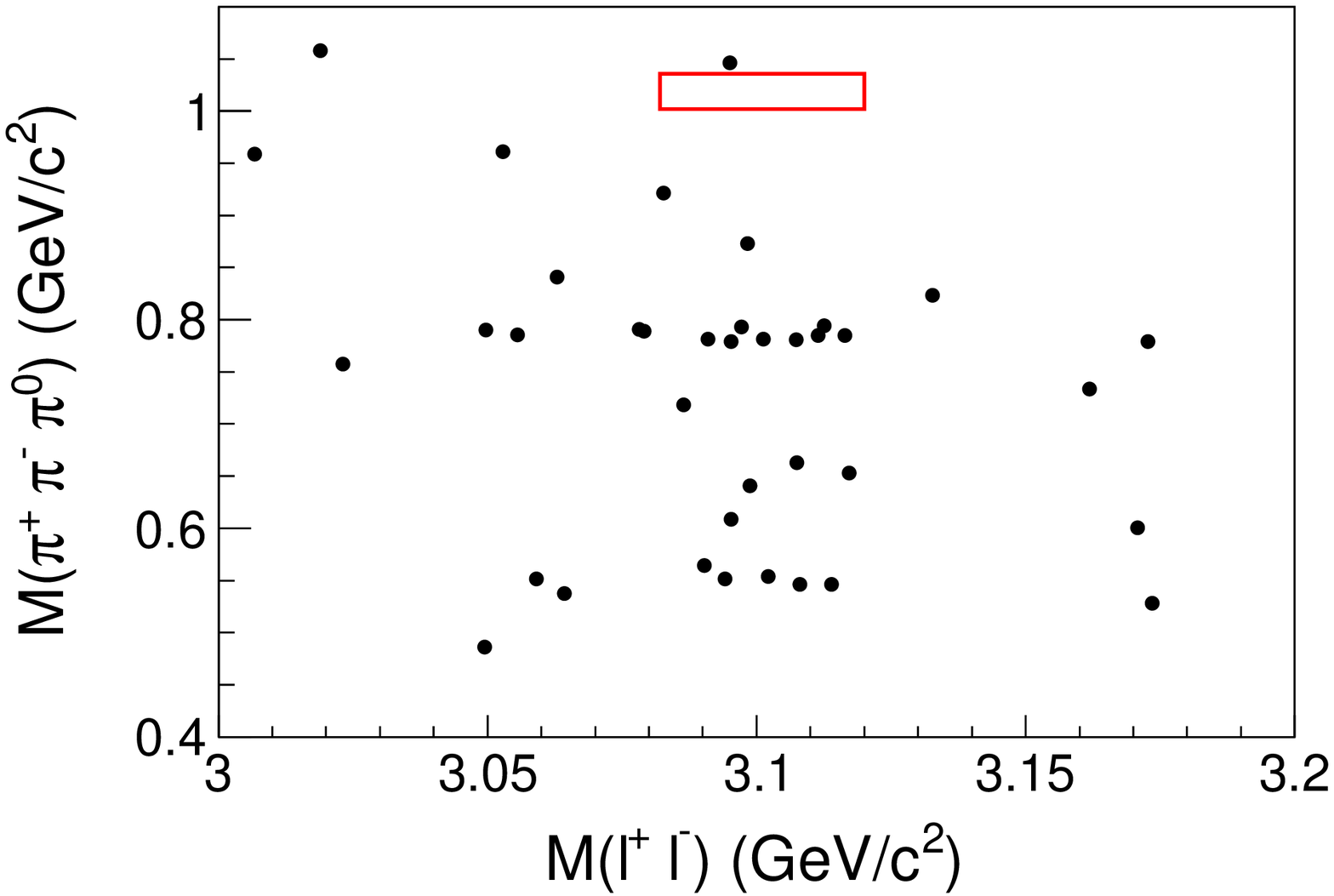}
\put(25,55){(b)}
\end{overpic}\\
\begin{overpic}[width=0.49\textwidth]{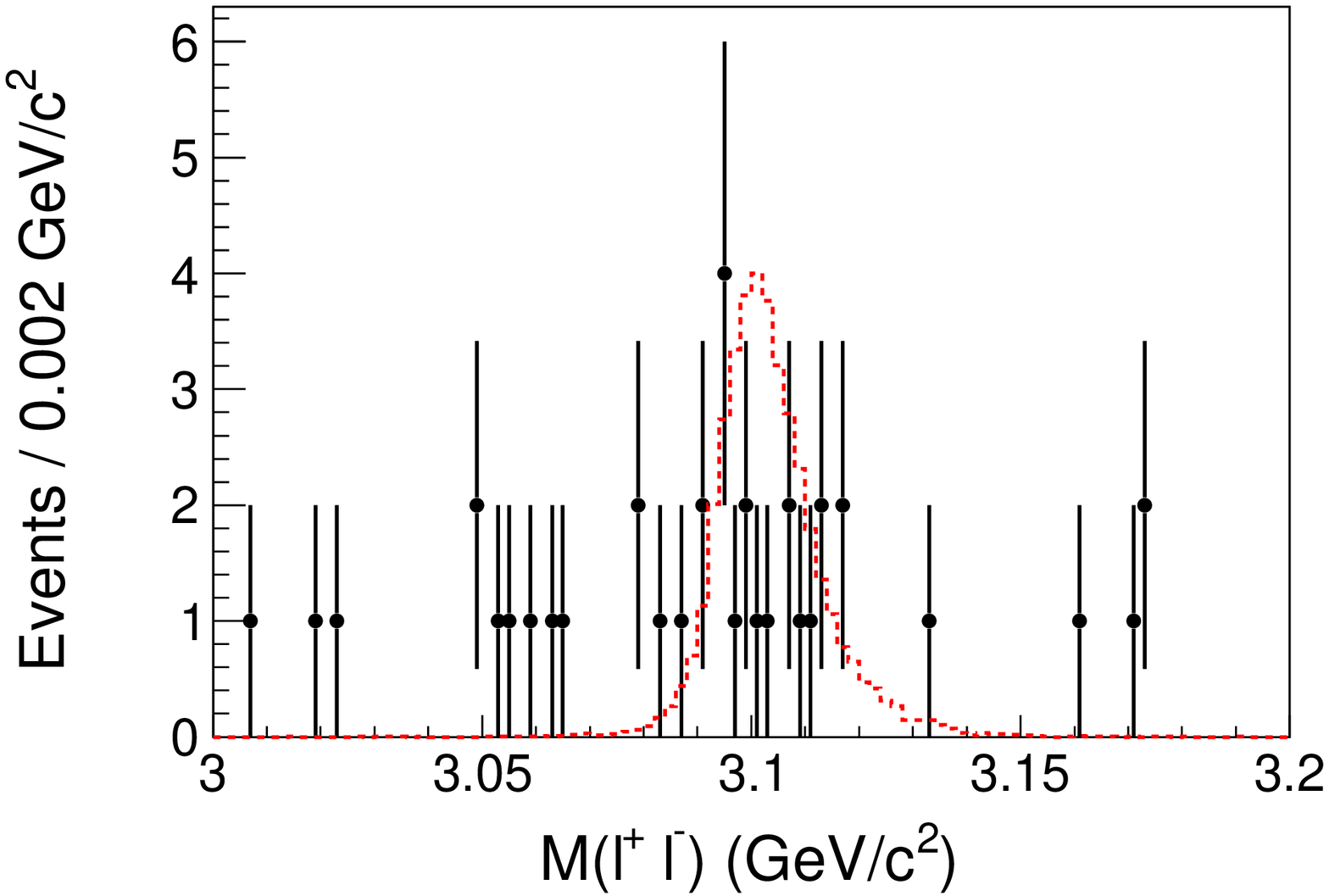}
\put(25,55){(c)}
\end{overpic}
\begin{overpic}[width=0.49\textwidth]{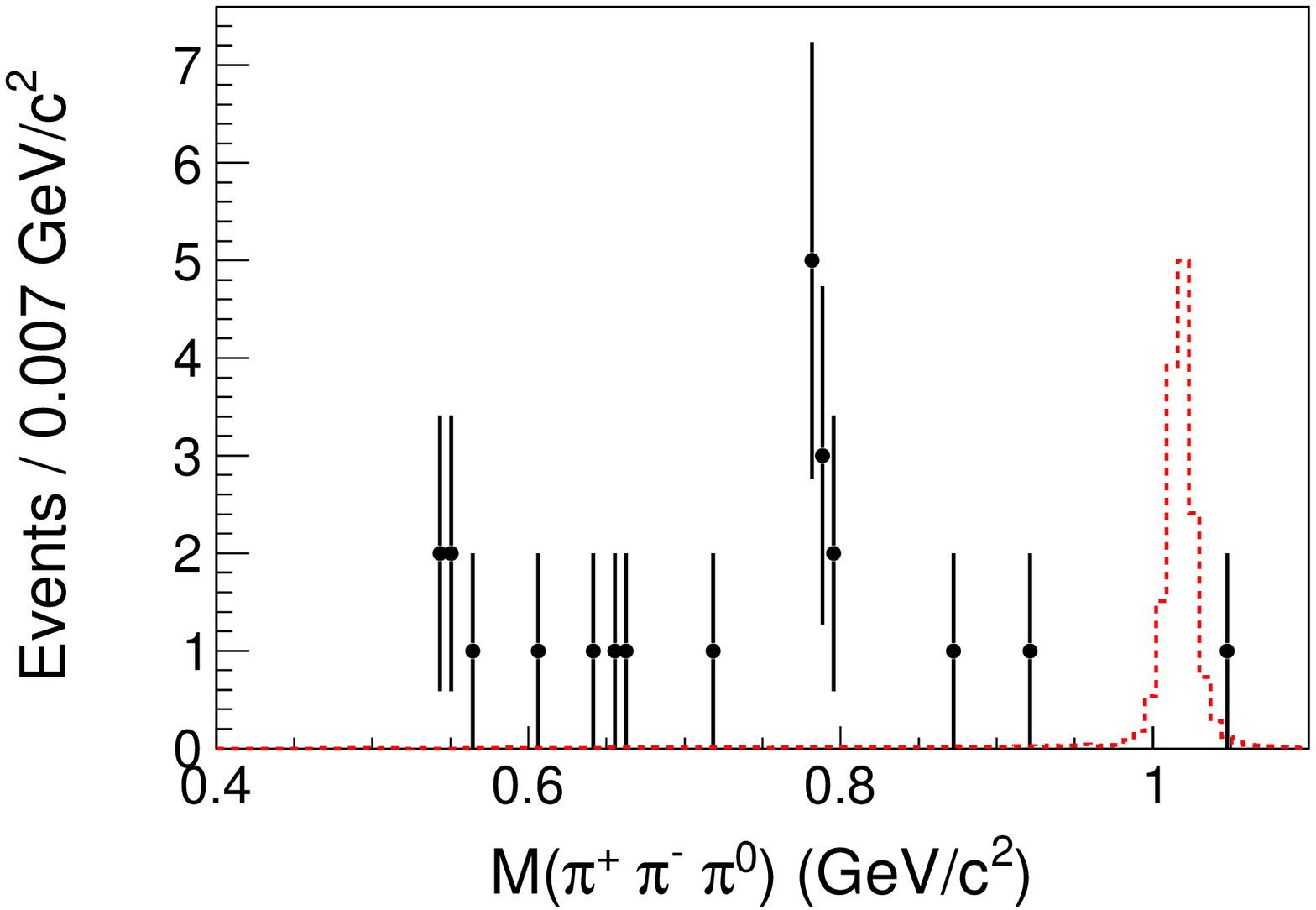}
\put(25,55){(d)}
\end{overpic}
\caption{Scatter plots for (a) signal MC, (b) data at 4.26 GeV, and
  the projections along (c) $M(\LL)$ and (d) $M(\pipipi)$. The red box
  shows the applied mass windows of $\phi$ and $\jpsi$. The red dashed
  histogram shows the MC simulated shape (with arbitrary normalization).}
\label{fig_scatter_pipipi}
\end{figure*}

Figure~\ref{fig_scatter_pipipi} shows the scatter plots of $M(\pipipi)$ vs.
$M(\LL)$ for MC and data at 4.26 GeV and the 1-D projections. The
dominant background events are from $\EE \to \omega\chi_{cJ}$ and $\EE \to
\eta \jpsi$ with a random photon. Neither of these channels can be
selected as $\gamma \phi \jpsi$ signal.

\begin{figure}[htbp]
\begin{center}
    \includegraphics[width=0.45\textwidth]{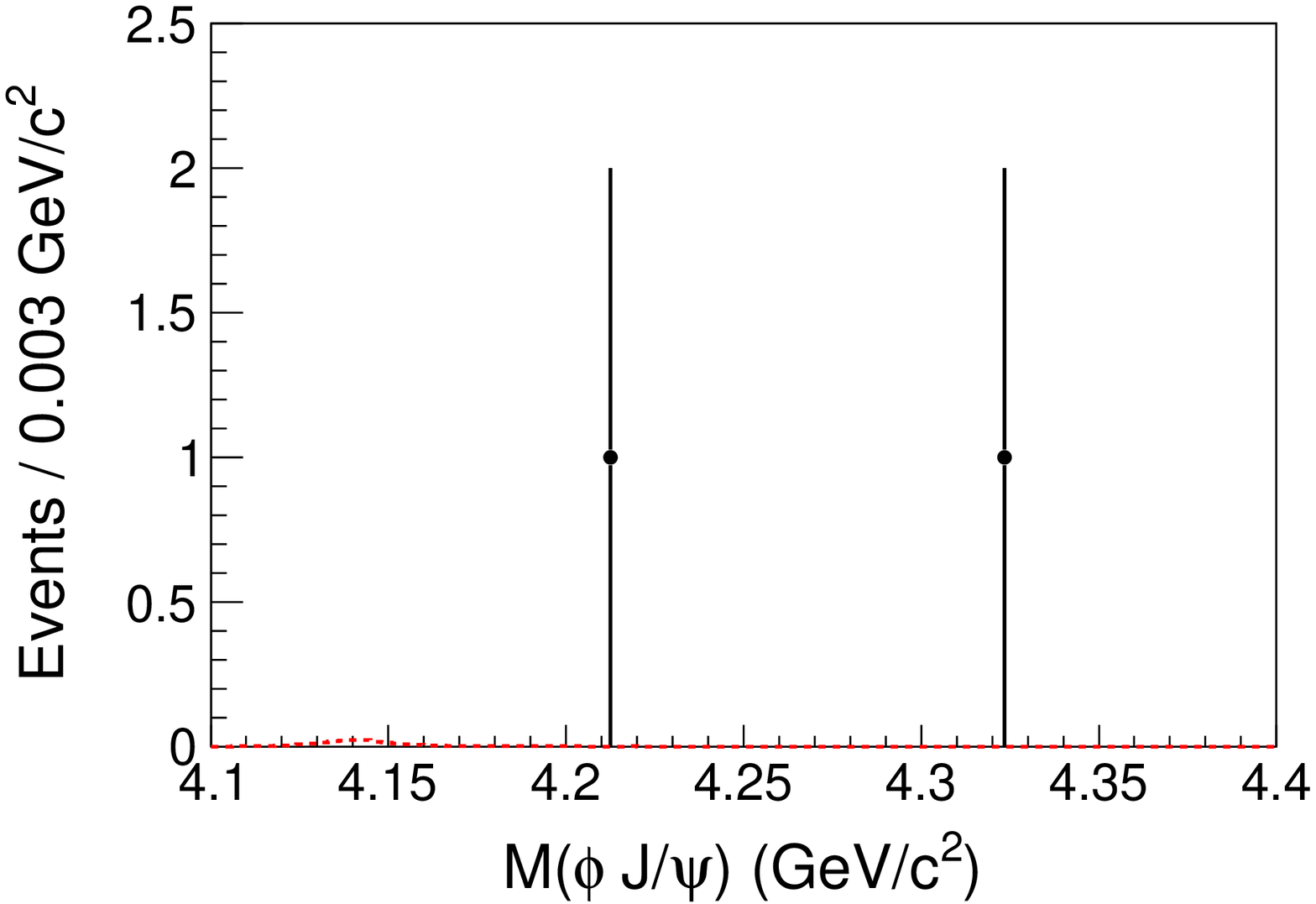}
\end{center}
\caption{Distribution of $M(\phi \jpsi)$ with $\phi \to \pipipi$
at $\sqrt{s}= 4.36$~GeV. The red dashed histogram represents the
signal MC events scaled to the measured upper limit.}
\label{fig:phijpsi_pipipi}
\end{figure}

From the inclusive MC sample and exclusive $\EE \to \pipipi \jpsi$
and $\eta \jpsi$ MC samples, correspond to production cross section of
200 pb, we find no events in the $Y(4140)$ signal region, so these
background channels are neglected. The production cross section
of the above two modes are at a few or a few tens of pb level~\cite{belle:etaj,cleo:kkj}. After the event selection,
there are no events left for the data samples at $\sqrt{s}= 4.23$
and $4.26$~GeV, and there are only two events left for the data
sample at $4.36$~GeV. Figure~\ref{fig:phijpsi_pipipi} shows the
distribution of $M(\phijpsi)=M(\pipipi \ell^+ \ell^-)- M(\ell^+
\ell^-)+m_{\jpsi}$ at $\sqrt{s}=4.36$~GeV.  Both surviving
events are far from the $Y(4140)$ signal region.

\begin{figure*}[htbp]
\begin{center}
\begin{overpic}[width=0.49\textwidth]{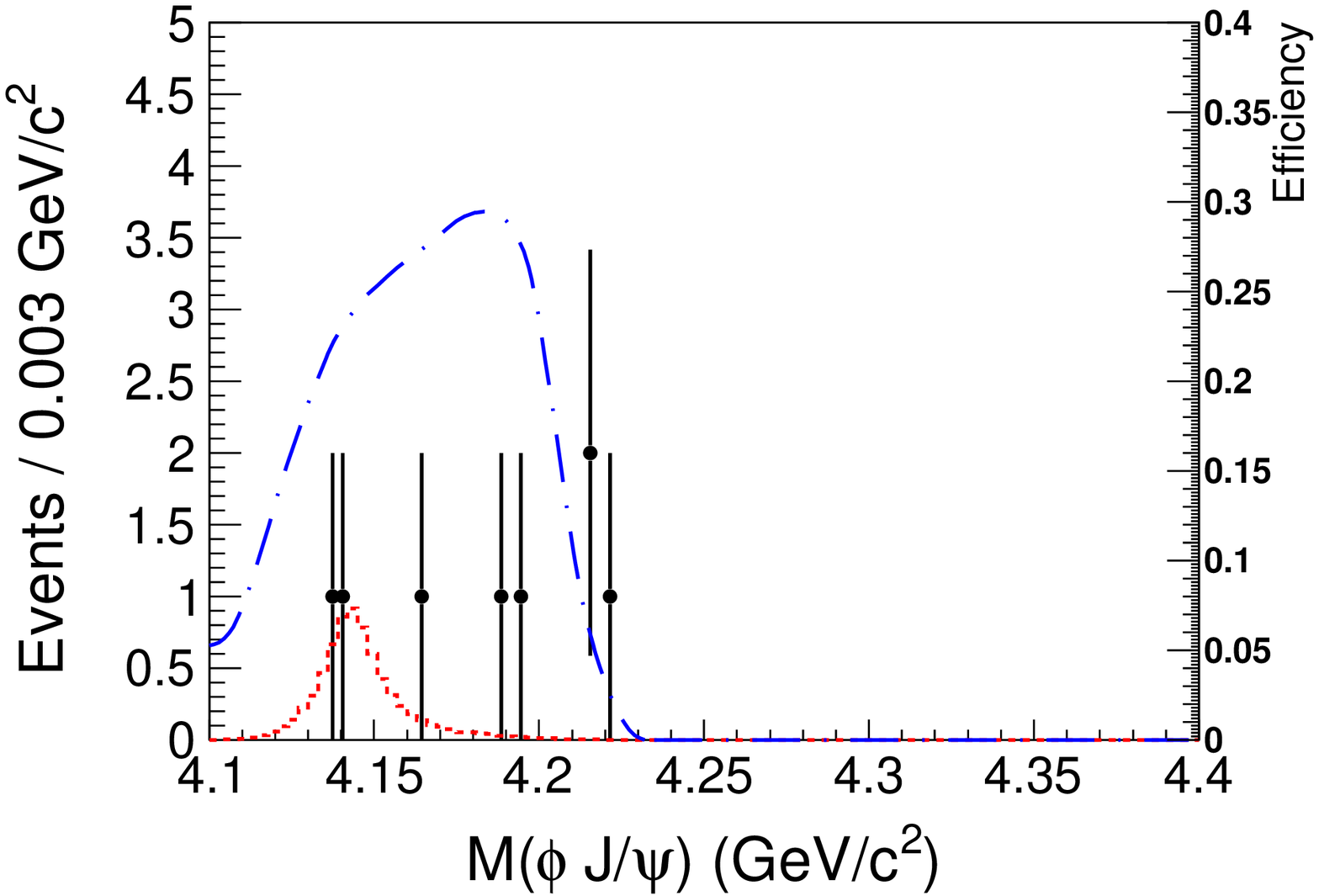}
\put(25,55){(a)}
\end{overpic}
\begin{overpic}[width=0.49\textwidth]{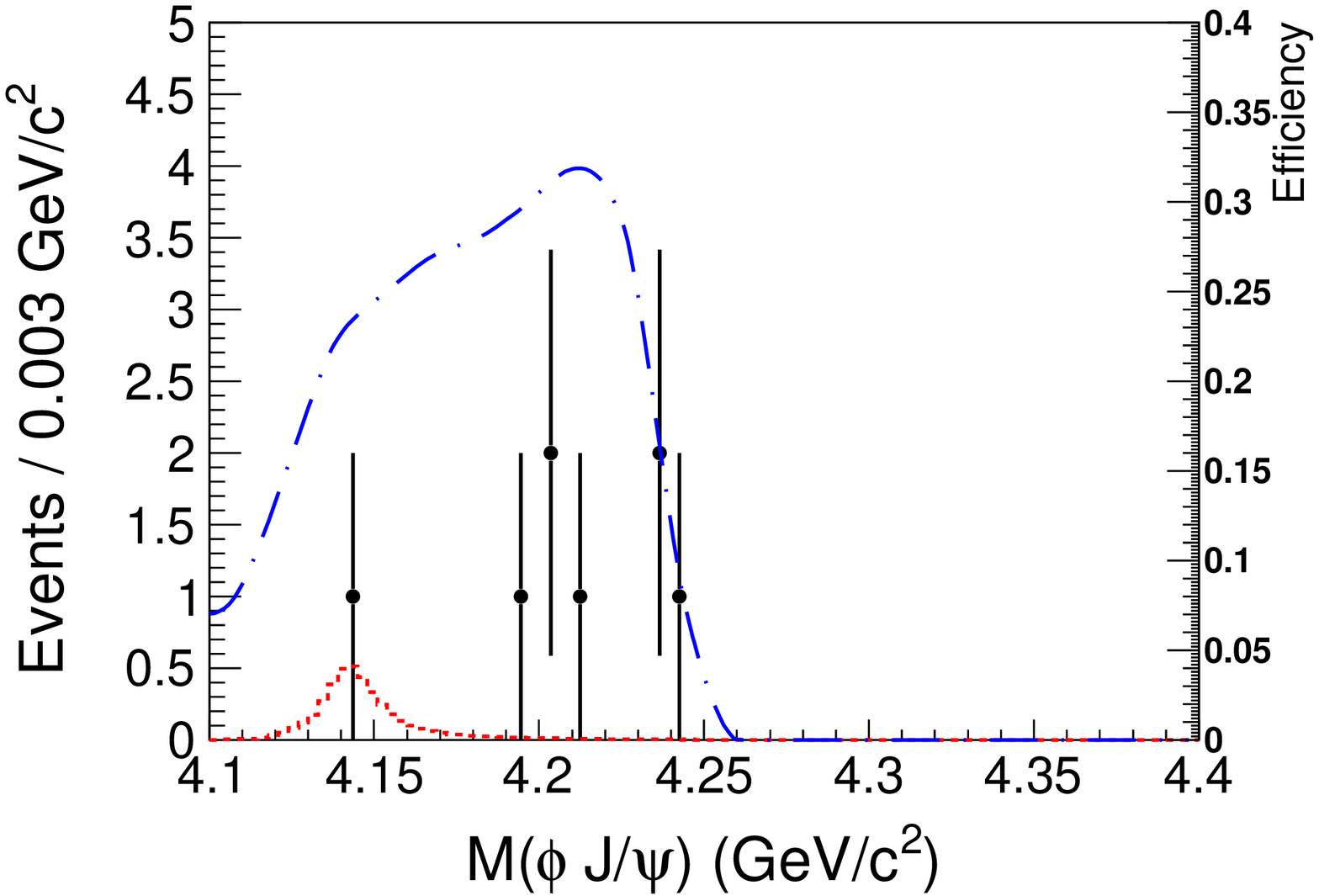}
\put(25,55){(b)}
\end{overpic}\\
\begin{overpic}[width=0.49\textwidth]{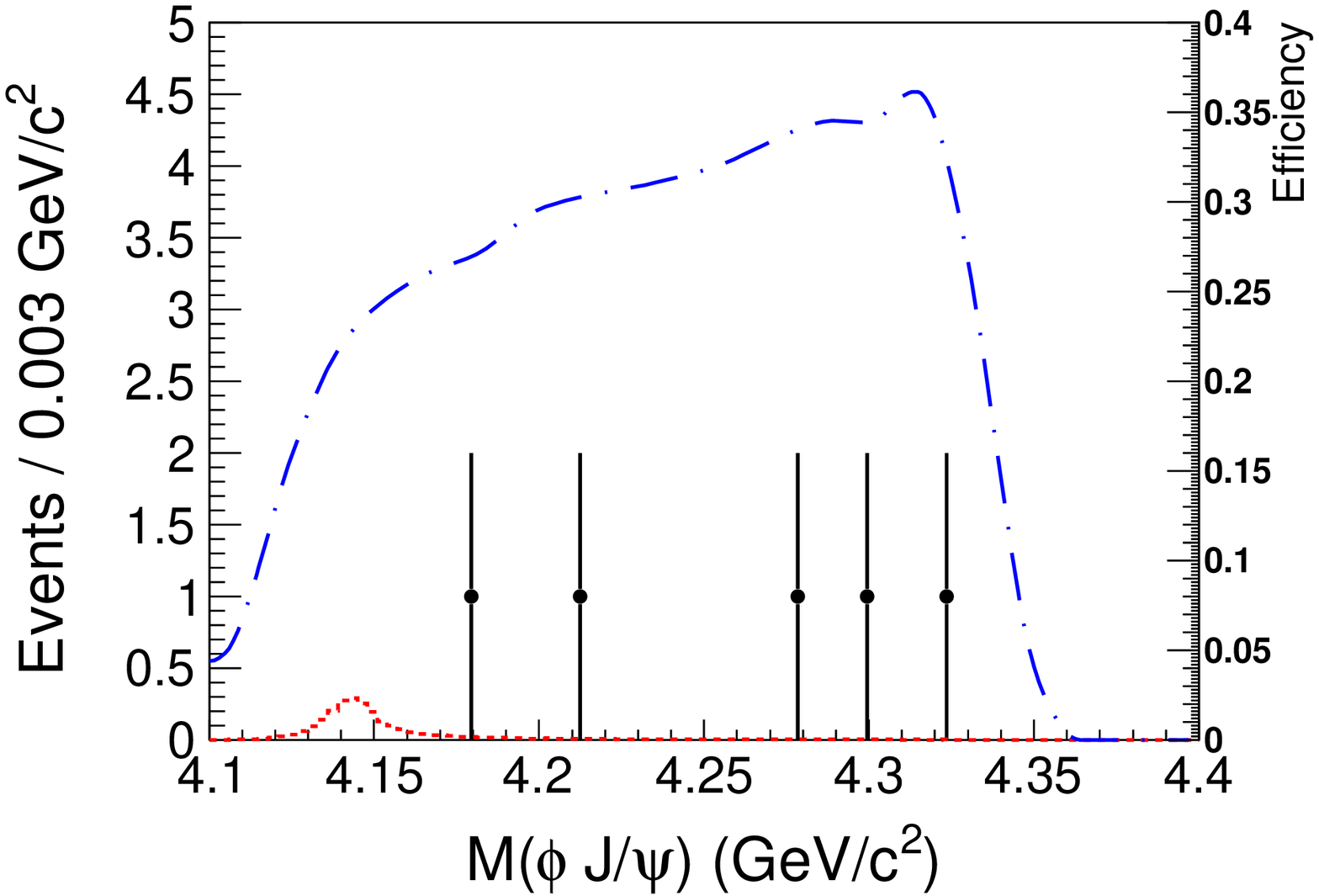}
\put(25,55){(c)}
\end{overpic}
\begin{overpic}[width=0.49\textwidth]{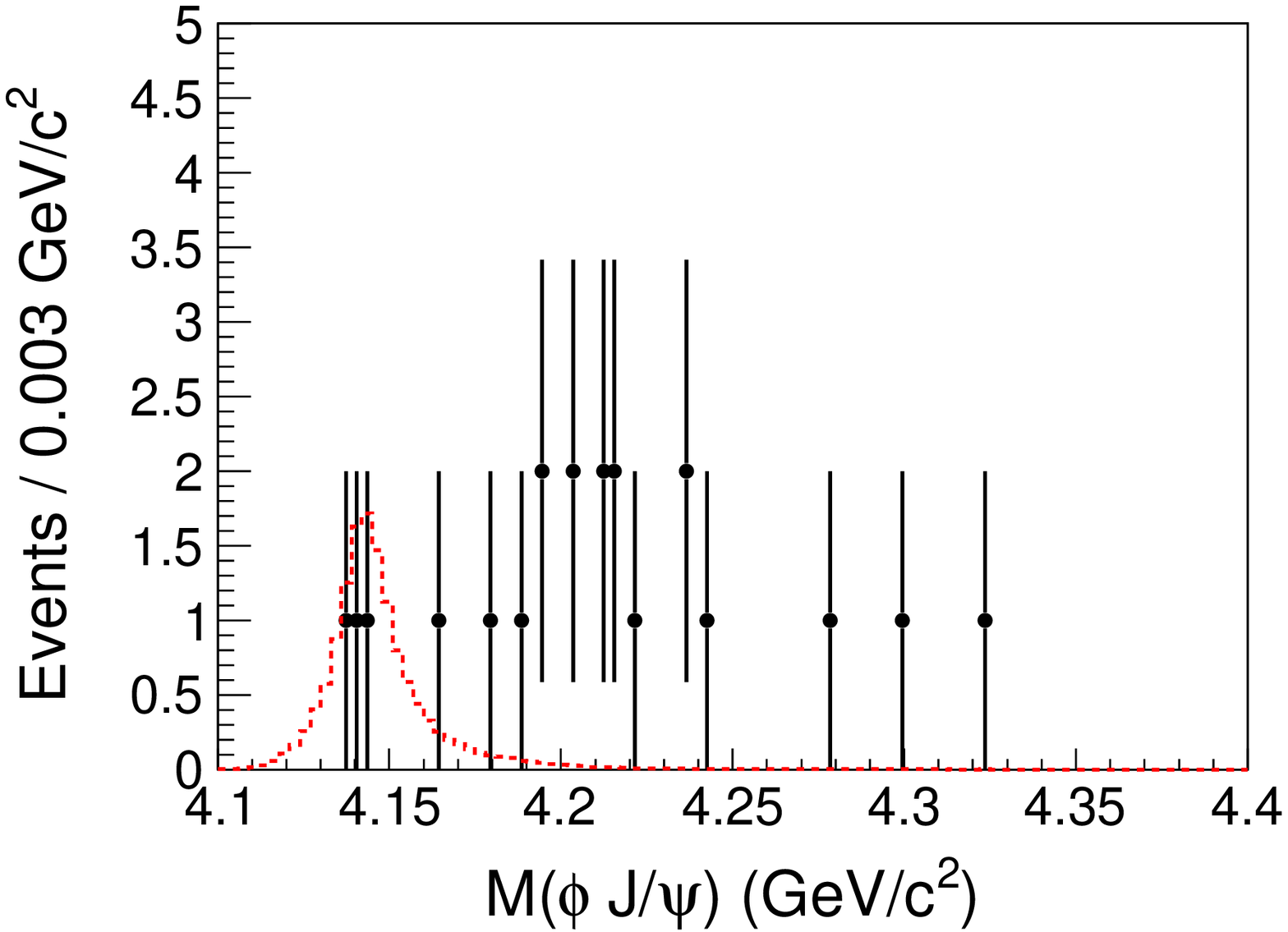}
\put(25,55){(d)}
\end{overpic}
\end{center}
\caption{Distribution of $M(\phijpsi)$ summed over all $\phi$ and
$\jpsi$ decay modes at $\sqrt{s}=$ (a) $4.23$, (b) $4.26$, (c)
$4.36$~GeV, and (d) the sum of three data samples. The red
dashed histogram  represents signal MC events scaled to our
measured upper limit. The blue dashed-dot line shows the efficiency
distribution.} \label{fig:phijpsi_sum1}
\end{figure*}

\section{Cross sections}
\label{sec:cross}

As the $Y(4140)$ signal is not significant, and it cannot be
distinguished from the contribution of the non-resonant processes due
to low statistics, we set an upper limit
on this production rate at the $90\%$ confidence level (C.L.). The
six decay modes (three $\phi$ modes $\times$ two $\jpsi$ modes)
are combined to obtain the best estimate of the $Y(4140)$ production
cross section by counting the numbers of events located in the
$Y(4140)$ signal region. This signal region is defined as
$M(\phi\jpsi) \in $[4.11, 4.17] GeV/$c^{2}$, which covers about
$95\%$ of the signal events according to the MC simulation. The
combined distributions of $M(\phi\jpsi)$ are shown in
Fig.~\ref{fig:phijpsi_sum1}. From MC studies of the known possible
background channels, which are detailed in Sec. III for
the three $\phi$ decay modes separately, no events in the signal region are observed.
Since information on possible backgrounds is limited, we
conservatively assume that all the events that lie in the signal region
are from the $Y(4140)$.  We assume that the number of observed
events follows Poisson distributions. The total likelihood of the
six modes is defined as
\begin{equation}
L(n^{\rm prod})=\prod_{i=1}^{6} P(N_{i}^{\rm obs};n^{\rm prod}\mathcal{B}_{i}\epsilon_{i}).
\end{equation}
Here $P(r;\mu)=\frac{1}{r!}\mu^{r} e^{-\mu}$ is the probability
density function of a Possion distribution, $n^{\rm prod}$ is the
number of produced $Y(4140)\to \phi \jpsi$ events, $N_{i}^{\rm
obs}$ is the number of observed events in the $i$th mode,
$\mathcal{B}_{i}$ and $\epsilon_{i}$ are the corresponding
branching fraction and efficiency, respectively. To take
systematic uncertainties into consideration, we convolute the
likelihood distribution with a Gaussian function with mean value
of $0$ and standard deviation $n^{\rm prod}\cdot\Delta$, where
$\Delta$ is the relative systematic uncertainty described in the
next section. The upper limit on $n^{\rm prod}$ at the $90\%$ C.L.
is obtained from $\int_{0}^{n^{\rm prod}}L(x)dx/
\int_{0}^{\infty}L(x)dx=0.9$.


The Born cross section is calculated using
\begin{equation}
\sigma^{B} = \frac{n^{\rm prod}}{\mathcal{L}_{\rm int}
(1+\delta)(1+\delta^{vac})},
\end{equation}
where $\mathcal{L}_{\rm int}$ is the integrated luminosity,
$(1+\delta)$ is the radiative correction factor, including initial
state radiation, $\EE$ self-energy and initial vertex correction,
and $(1+\delta^{vac})$ is the vacuum polarization factor,
including leptonic and hadronic parts.

The radiative correction factor ($1+\delta$) is obtained by using
a QED calculation~\cite{qed_radi}. We assume that the cross
section for $\EE \rightarrow \gamma Y(4140)$  follows the $Y(4260)
\rightarrow \pp \jpsi$ line shape, and use the Breit-Wigner
parameters of the $Y(4260)$~\cite{pdg} as input. The values for
$(1+\delta)$ are listed in Table~\ref{tab:cross_section}. The
vacuum polarization factor ($1+\delta^{vac}$)=1.054 is taken from
Ref.~\cite{vacuum:polarization}, and its uncertainty in comparison
with other uncertainties is negligible.

The upper limit on $\sigma^{B}$ is obtained by replacing $n^{\rm
prod}$ with the upper limit on $n^{\rm prod}$. The upper limits on
the product of the Born cross section and branching fraction
$\sigma[\EE \rightarrow \gamma Y(4140)] \cdot \mathcal{B}(Y(4140)
\rightarrow \phijpsi)$ at the $90\%$ C.L. are 0.35, 0.28 and
0.33~pb for $\sqrt{s}=$ 4.23, 4.26 and 4.36~GeV, respectively. The
results are listed in Table~\ref{tab:cross_section}.

\begin{table}[htbp]
\caption{Upper limits at the $90\%$ C.L. for measurements of
$\sigma^{B} \cdot \mathcal{B}=\sigma(\EE \rightarrow \gamma
Y(4140)) \cdot \mathcal{B}(Y(4140) \rightarrow \phijpsi)$. }
\label{tab:cross_section}
\begin{tabular}{c | c | c | c | c}
\hline \hline
$\sqrt{s}$ (GeV) & Luminosity (pb$^{-1}$)& $(1+\delta)$ & $n^{\rm prod}$ & \quad $\sigma^{B} \cdot \mathcal{B}$ (pb)\\
\hline
4.23 & 1094 &0.840 & $<339$ & $<0.35$ \\
4.26 & 827 &0.847  & $<207$ & $<0.28$\\
4.36 & 545 &0.944  & $<179$ & $<0.33$ \\
\hline
\hline
\end{tabular}
\end{table}


\section{Systematic uncertainties}
\label{sec:syst}

The sources of the systematic uncertainties are listed in
Table~\ref{tab:sys_err} for the measurement at $4.26$~GeV and are
explained below.

\begin{table}[htbp]
\caption{Summary of systematic uncertainties for
$\sqrt{s}=$4.26~GeV data sample.} \label{tab:sys_err}
\begin{tabular}{c c c c}
\hline \hline
    \multirow{2}{0.01\textwidth}{\centering{Source}} & \multicolumn{3}{c}{Systematic uncertainty (\%)} \\
    {} & $\phi \rightarrow \kk$ & $ \KK$ & $\pipipi$ \\
    \hline
    Luminosity & 1.0 & 1.0 & 1.0 \\
    Tracking & 3.0 & 2.0 & 4.0 \\
    Photon & 1.0 & 1.0 & 3.0 \\
    PID & 1.0 & - & - \\
    $\ks$ reconstruction & - & 4.0 & - \\
    Branching fraction & 1.2 & 1.3 & 2.2 \\
    Radiative correction & 3.8 & 3.8 & 3.8 \\
    Radiative decay & \multirow{2}{15pt}{11.5} &\multirow{2}{12pt}{8.8} &\multirow{2}{15pt}{13.5} \\
    distribution &  &  & \\
   Kinematic fit & 3.8 & 6.4 & 3.2 \\
    \hline
    Total & 13.2 & 12.5 & 15.4 \\
    \hline \hline
\end{tabular}
\end{table}
The luminosity is measured using Bhabha events, with an
uncertainty less than 1.0\%~\cite{x3872:bes3}. The difference between
data and MC in tracking efficiencies for charged tracks
is $1.0\%$ per track~\cite{pidds:bes3}. Studies with a
sample of $\jpsi \to \rho \pi$ events show that the uncertainty in
the reconstruction efficiency for photons is less than
1.0\%~\cite{photon:bes3}.  For the $\phi \to \kk$ mode, PID is
required for the kaons, and this is taken as
$1.0\%$~\cite{pidds:bes3} per track. Since we require only one
kaon to be identified, the uncertainty is smaller than $1.0\%$,
but we take $1.0\%$ to be conservative. For the $\ks$
reconstruction, the difference between data and MC simulation is
estimated to be 4.0\% including tracking efficiencies for two
daughter pions from the study of $\jpsi \to K^{*} \bar{K}^{0} +
c.c.$~\cite{ks_recons}.

The branching fractions for $\phi\to \kk$, $\KK$ and $\pipipi$,
and $\jpsi \to \EE$ and $\MM$ are taken from the PDG~\cite{pdg}.
The uncertainties of the branching fractions are taken as
systematic uncertainties, which are 1.2\%, 1.3\%, and 2.2\% for
the process with $\phi \to \kk$, $\KK$, and $\pipipi$,
respectively.

The radiative correction factor and detection efficiency are
determined under the assumption that the production $\EE \to
\gamma Y(4140)$ follows the $Y(4260)$ line shape. The $Y(4360)$
line shape~\cite{pdg} is used as an alternative assumption, and
the difference in $\epsilon\cdot(1+\delta)$ is taken as a
systematic uncertainty. This is $3.3\%, 3.8\%$, and $10.0\%$ for
$\sqrt{s}= 4.23, 4.26$, and $4.36$~GeV, respectively; the value
for $\sqrt{s}= 4.36$~GeV is larger than others, since the
line shape changes the biggest at this energy point.

The $J^{P}$ of the $Y(4140)$ is unknown, and the efficiency is
obtained from a MC sample generated uniformly in phase space. In
order to estimate the uncertainty due to decay dynamics, the
angular distribution of the radiative photon is generated as $1+
\cos^{2}\theta$ and $1- \cos^{2}\theta$ to determine the
difference of efficiency from that of the phase space MC sample.
We take the biggest difference as the systematic uncertainty of
the radiative decay distribution, which is $11.5\%$, $8.8\%$, and
$13.5\%$ for the modes $\phi \to \kk$, $\KK$, and $\pipipi$,
respectively.

For the $\jpsi$, $\phi$, $\ks$ and $\pi^{0}$ mass windows, the
selection is very loose, so the difference between data and MC
simulation samples are negligible.

For the uncertainties due to kinematic fitting and vertex fitting,
it is hard to find an appropriate control sample to measure them.
A correction to the track helix parameters in the MC
simulation~\cite{guo_pull} was applied so that the distribution of
the MC simulation events is similar to that of the data, and
we take half of the difference between the efficiency with and
without this correction as the systematic uncertainty.  The MC sample
with the track helix parameter correction applied is used as the default
in this analysis.

Assuming that all sources of systematic uncertainties are
independent, the total errors are given by the quadratic sums of
all of the above. At $4.26$~GeV, the values, which are listed in Table~\ref{tab:sys_err}, are $13.2\%$,
$12.5\%$, and $15.4\%$, for the modes $\phi \to \kk$, $\KK$, and
$\pipipi$, respectively. For the events collected at $4.23$ and
$4.36$~GeV, the only difference is the systematic uncertainty due
to $(1+\delta)$, and the total systematic errors are $13.1\%$,
$12.4\%$, and $15.3\%$ for events at $4.23$~GeV, and $16.1\%$,
$15.4\%$, and $17.9\%$, for events at $4.36$~GeV.

\section{Results and discussions}
\label{sec:summ}

In summary, we search for the $Y(4140)$ via $\EE \to
\gamma \phijpsi$ at $\sqrt{s} = 4.23, 4.26$, and $4.36$~GeV
and observe no significant $Y(4140)$ signal in either data sample.  The upper
limits of the product of cross section and branching fraction
$\sigma[\EE \rightarrow \gamma Y(4140)] \cdot \mathcal{B}(Y(4140)
\rightarrow \phijpsi)$ at the $90\%$ C.L. are estimated as $0.35,
0.28$, and $0.33$~pb at $\sqrt{s} = 4.23, 4.26$, and $4.36$~GeV,
respectively.

These upper limits can be compared with the $X(3872)$ production
rates~\cite{x3872:bes3}, which were measured with the same data
samples by BESIII. The latter are $\sigma[\EE \to  \gamma X(3872)]
\cdot \mathcal{B}(X(3872) \rightarrow  \pi^{+} \pi^{-} \jpsi) =
[0.27 \pm 0.09({\rm stat}) \pm 0.02({\rm syst})]$~pb, $[0.33 \pm
0.12({\rm stat}) \pm 0.02({\rm syst})]$~pb, and $[0.11 \pm
0.09({\rm stat}) \pm 0.01({\rm syst})]$~pb at $\sqrt{s} = 4.23,
4.26$, and $4.36$~GeV, respectively, which are of the same order
of magnitude as the upper limits of $\sigma[\EE \rightarrow \gamma
Y(4140)] \cdot \mathcal{B}(Y(4140) \rightarrow \phijpsi)$ at the
same energy.

The branching fraction $\mathcal{B}(Y(4140)\to \phi \jpsi)$ has
not previously been measured. Using the partial width of $Y(4140) \to
\phi\jpsi$ calculated under the molecule
hypothesis~\cite{y4140exp:4}, and the total width of the $Y(4140)$
measured by CDF~\cite{y4140b}, the branching fraction is estimated
roughly to be 30\%. A rough estimation for $\mathcal{B}(X(3872)
\rightarrow \pi^{+} \pi^{-} \jpsi)$ is $5$\%~\cite{x3872b}.
Combining these numbers, we estimate the ratio ${\sigma[\EE
\to\gamma Y(4140)]}/{\sigma[\EE \to\gamma X(3872)]}$ is at the
order of 0.1 or even smaller at $\sqrt{s} = 4.23$ and 4.26~GeV.

\acknowledgements

The BESIII collaboration thanks the staff of BEPCII and the IHEP
computing center for their strong support. This work is supported in
part by National Key Basic Research Program of China under Contract
No.~2015CB856700; Joint Funds of the National Natural Science
Foundation of China under Contracts Nos.~11079008, 11179007, U1232201,
U1332201; National Natural Science Foundation of China (NSFC) under
Contracts Nos.~10935007, 11121092, 11125525, 11235011, 11322544,
11335008; the Chinese Academy of Sciences (CAS) Large-Scale Scientific
Facility Program; CAS under Contracts Nos.~KJCX2-YW-N29, KJCX2-YW-N45;
100 Talents Program of CAS; INPAC and Shanghai Key Laboratory for
Particle Physics and Cosmology; German Research Foundation DFG under
Contract No.~Collaborative Research Center CRC-1044; Istituto
Nazionale di Fisica Nucleare, Italy; Ministry of Development of Turkey
under Contract No.~DPT2006K-120470; Russian Foundation for Basic
Research under Contract No.~14-07-91152; U.S.\ Department of Energy
under Contracts Nos.~DE-FG02-04ER41291, DE-FG02-05ER41374,
DE-FG02-94ER40823, DESC0010118; U.S.~National Science Foundation;
University of Groningen (RuG) and the Helmholtzzentrum fuer
Schwerionenforschung GmbH (GSI), Darmstadt; WCU Program of National
Research Foundation of Korea under Contract No.~R32-2008-000-10155-0.

\end{spacing}

\end{document}